\documentclass[fleqn,12pt]{article}

\usepackage[T1]{fontenc}
\usepackage[utf8]{inputenc}

\usepackage{caption}
\usepackage{subcaption}
\usepackage{amsmath}
\usepackage{float}
\usepackage[export]{adjustbox}
\usepackage{array}

\usepackage[numbers]{natbib}

\usepackage{ntheorem}

\usepackage{authblk}
\usepackage{url}
\usepackage{booktabs}

\usepackage[
  left=20mm,
  right=20mm,
  top=20mm,
  bottom=25mm
]{geometry}

\usepackage{setspace}
\title{{\bf Stock Price Responses to Firm-Level News in Supply Chain Networks}}

\author[1,2]{Hiroyasu Inoue}
\author[3,4]{Yasuyuki Todo}

\affil[1]{University of Hyogo, Graduate School of Information Science, Kobe 6500047, Japan}
\affil[2]{RIKEN, Center for Computational Science, Kobe 6500047, Japan}
\affil[3]{Waseda University, Graduate School of Economics, Tokyo 1698050, Japan}
\affil[4]{Research Institute of Economy, Trade and Industry, Tokyo 1008901, Japan}

\affil[]{\protect\url{inoue@gsis.u-hyogo.ac.jp}}

\date{}

\begin{document}
\maketitle

\begin{abstract}
\noindent
This study examines how positive and negative news about firms is associated with stock prices and whether these associations extend to firms' suppliers and clients connected through supply chain relationships, using large samples of publicly listed firms worldwide and in Japan. News sentiment is measured using FinBERT, a natural language processing model fine-tuned for financial texts, and supply chain links are identified from financial statements for global firms and from large-scale firm-level surveys for Japanese firms. We find that stock prices exhibit systematic associations with positive and negative news even before public disclosure. These associations are also observed for suppliers and clients before and after disclosure. In general, relative to the pre-disclosure period, post-disclosure associations are larger, with the difference concentrated around the time of public news disclosure. However, for Japanese firms, the post-disclosure associations for suppliers and clients are smaller than the pre-disclosure associations, in contrast to the pattern observed for firms outside Japan.
\end{abstract}

\flushbottom

\thispagestyle{empty}

\newpage

\section{Introduction} 

Firms interact and influence one another through various types of networks. Major firm networks that play a significant role in firms' behavior and performance in the literature include financial networks through ownership \cite{acemoglu2015systemic,elliott2014financial,gai2010contagion} and knowledge networks through research collaboration \cite{hansen2002knowledge,grant1995knowledge,brusoni2001knowledge,todo2016strength,iino2021}. Another type of network that has received increasing attention in the academic literature and business and policy fields is supply chains formed through transactions in materials, parts, and components \cite{coe2008global,henderson2002global}. One reason for the growing attention is that supply chains have expanded globally, linking firms worldwide within a small number of steps to one another \cite{baldwin2016supplychains}.

Global supply chains have rapidly expanded because firms interconnected through supply chains can greatly benefit from the network. Most notably, supply chains can empower firms by enhancing the efficiency of production processes and thus improving firms' economic performance \cite{cao2011supply,baldwin2013}. For instance, by procuring and manufacturing different goods in various locations based on each location's comparative advantages and cost-effectiveness, firms can streamline their operations as a network, ultimately leading to reductions in production costs and improvements in productivity \cite{dainty2001new}. Moreover, it is frequently observed that productivity spills over through supply chains from upstream suppliers to downstream client firms because of high-quality materials, parts, components, and services, and from downstream to upstream firms because of the learning and diffusion of knowledge and technology \cite{mundt2023peer,todo2016strength,javorcik2004}.

Conversely, supply chains can be a channel of negative shocks that adversely affect firms' performance, particularly once supply chains are disrupted due to unexpected events, such as natural disasters and geopolitical instability \cite{boehm2019input,inoue2019firm,barrot2016input}. Such negative shocks examined in the literature include the Great East Japan Earthquake in 2011 \cite{matsuo2015implications}, the massive flood in Thailand in 2011 \cite{haraguchi2015flood}, the COVID-19 pandemic from 2020 to 2022 \cite{cui2023exploring}, and the Russia-Ukraine war that started in 2022 \cite{moosavi2022supply}.

Accordingly, prior studies show that positive or negative shocks propagate through supply chains, affecting the performance of firms linked through supply chains \citep{barrot2016input}. Because the corporate value of publicly listed firms is assessed in the stock market \cite{mitchell1994impact,tetlock2011all,chan2003stock}, a shock to a firm is expected to affect not only its own stock price but also the stock prices of its suppliers and clients. For example, following the Boeing 737 MAX crashes in 2019 and 2020, the stock prices of Boeing, some of its suppliers such as GE and Allegheny Technologies, and some of its clients such as American Airlines and Southwest Airlines declined sharply \cite{cnbc2019, cnbc20192, economists2019, marketwatch2019, reuters2019}. Also, misconduct by Daihatsu Motor Co., a Japanese automobile manufacturer affiliated with Toyota, in 2023 was accompanied by stock price declines for Daihatsu, Toyota, and related firms including suppliers and clients such as Aisin \cite{reuters20232, reuters2023}.

These examples show that shocks can propagate through supply chains and affect the stock prices of firms connected through supply chains. Although a substantial body of previous studies has examined the direct impact of shocks to firms on their own stock prices \cite{li2014news,sousa2019bert,souma2019enhanced}, there are relatively few studies that have investigated the diffusion of shocks to other firms' stock prices through supply chains.
\citet{cohen2008economic} show that information about a firm’s customers predicts future stock returns of its suppliers, providing early evidence that stock price reactions propagate through supply chain relationships.
More recently, \citet{tran2023esg} show that firm-level news shocks, particularly ESG-related news,
affect not only the focal firm’s stock returns but also the returns of its suppliers and clients,
providing direct evidence of news-driven spillovers along supply chain relationships.
Complementing this evidence, \citet{agarwal2017cosearch} demonstrate that investor attention,
measured through online co-search behavior, facilitates information diffusion across supply chain
partners and contributes to predictable stock return comovements.
Kim and Wagner \cite{kim2021examining} analyzed 315 cases reported by the Wall Street Journal regarding supply chain risks, which also included mentions of 69 suppliers and 246 clients.
Pandit et al. used a larger dataset of quarterly earnings announcements, which differ from the news reports considered in our study, and analyzed their propagation across 1,518 suppliers and 4,010 clients in COMPUSTAT \cite{pandit2011information}.
In addition, similar studies focusing on China are discussed \cite{li2021information,liu2023effects}.
However, existing studies remain constrained in two important respects.
First, most analyses focus on stock price reactions observed after news disclosures or identifiable events, leaving stock price movements prior to disclosure largely unexplored.
Second, evidence based on large-scale data covering long time horizons and extensive supply-chain networks remains scarce.

To fill this gap, this study examines whether news sentiment is associated with stock price changes of firms and their suppliers and clients in the pre- and post-disclosure periods, extending the analysis from market- or firm-level settings to supply-chain networks and cross-country comparisons. Using two large firm-level samples, one covering most publicly listed firms worldwide and the other focusing on Japanese firms, we test whether the level of positive and negative sentiment about a firm in a news article is associated with changes in the stock prices of the firm and its suppliers and clients before and after news disclosure. Our analysis combines more than 20 million news articles from the Thomson Reuters News Archive, comprehensive firm-level linkages that identify detailed supply chains based on financial statements, news articles, websites, and firm-level surveys, and daily stock price data for all listed firms, enabling an examination of both short- and long-run associations.

In this study, we adopt a reduced-form perspective and examine the relationship between news sentiment and stock prices without explicitly specifying investors’ information processing or decision-making. This abstraction includes how investors perceive and incorporate supply-chain relationships into their trading behavior.

\section{Data}

\subsection{News Articles}

We use news articles written in English from the Thomson Reuters News Archive for the period 2003--2016. In the News Archive, multiple news articles may be filed for a single event. Typically, the first article is an alert, which is a short message containing only essential facts. This is followed by a newsbreak, usually released within 5--20 minutes after the alert. A newsbreak consists of a headline, which is similar to that of the alert, and two to four paragraphs of body text.
Subsequent updates may follow the newsbreak after an additional 20--30 minutes, providing further information.

In this study, we use the newsbreak corresponding to each event. Although we are interested in the timing of public disclosure, alerts often do not include sufficient details about the related firms, whereas newsbreaks do. The time difference between alerts and newsbreaks is typically less than 20 minutes, which is negligible given that our analysis is conducted at the daily frequency. Accordingly, the release time of the newsbreak is aligned with the next market closing time of the relevant stock market, and the corresponding trading day is defined as day zero in the subsequent analyses. The newsbreak includes the time, headline, main text, Refinitiv Identification Codes (RICs) of one or more listed firms mentioned in the article, and other metadata.

\begin{figure}[tb]
\centering
\includegraphics[width=.35\linewidth]{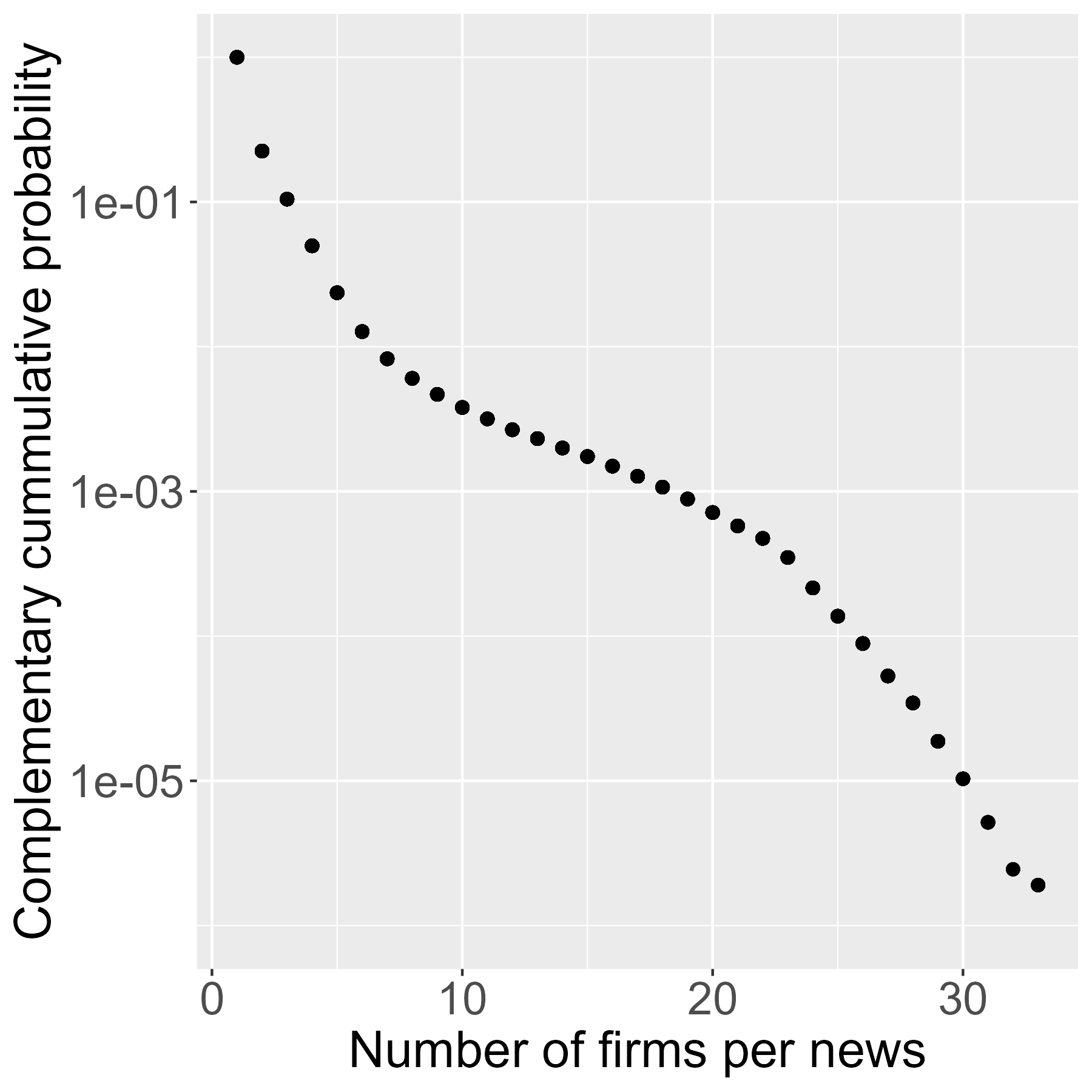}
\caption{Distribution of the number of firms mentioned in each news article.}
\label{fig:firmpernews}
\end{figure}

\begin{figure}[tb]
\centering
\includegraphics[width=.35\linewidth]{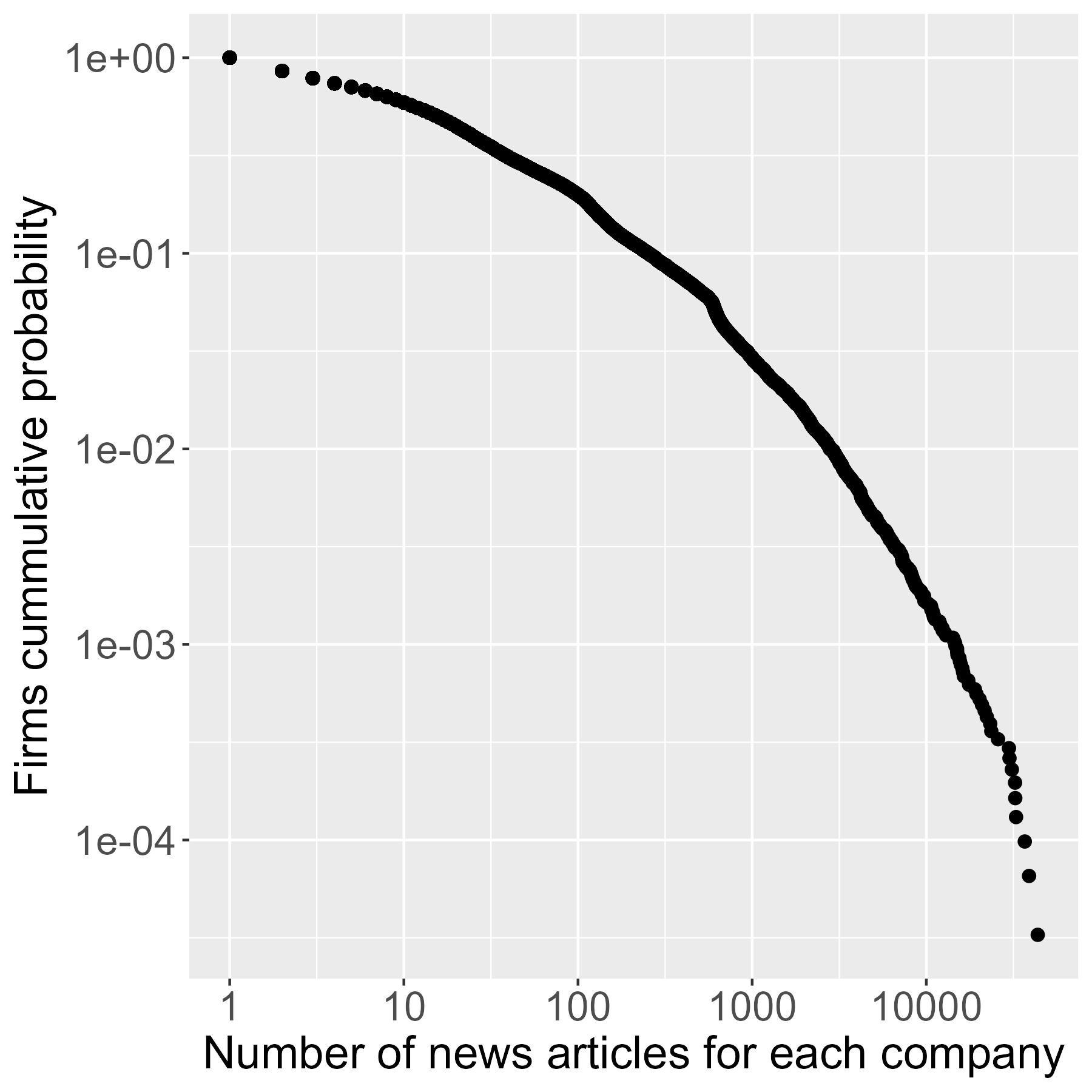}
\caption{Distribution of the number of news articles that mentioned each firm during the period 2003-2016.}
\label{fig:newsperfirm}
\end{figure}

As we will explain in detail later, our sentiment analysis, which determines the degree of positivity or negativity of each news article, utilizes the main text. The average number of words in an article is 243, whereas its median is 136 and its maximum is 3,911. Although the total number of news articles in the News Archive for the period examined is 20,803,561, we focus on 3,447,425 that mention any publicly listed firm in 105 stock markets in the world. Among these news articles, 90\% mention three or fewer firms, while the maximum number of firms mentioned in a single article is 33 (Figure \ref{fig:firmpernews}). The number of news articles that mention a particular listed firm during the period 2003--2016 is highly skewed. More than half of firms (55.6\%) were never mentioned, while 0.1\% were mentioned more than 10,000 times (Figure \ref{fig:newsperfirm}).


\subsection*{Stock Prices}

We utilize daily stock prices of firms that are currently listed in global stock markets,
obtained from the Refinitiv Eikon database, one of the largest providers
of financial markets data and infrastructure, through its Application Programming Interface (API).
Firms that were listed in the past but are no longer listed are excluded from the sample,
as their stock price data are not available through the API.
We define the daily stock price of each firm as its closing price.
Accordingly, the stock price data include RICs,
dates, and closing prices.

In addition to stock prices of each firm, we utilize the Refinitiv index for each stock market that indicates the average stock price level in the market to control for the market-specific time trend. The Refinitiv index is available for 105 major markets that cover 92\% of listed firms worldwide. 

\subsection*{Supply Chains}

To identify global supply chains among publicly listed firms in the world, we rely on data from FactSet that cover a number of firms worldwide for the period 2003--2016, as of January 18, 2023. The FactSet data include major clients of each firm reported in its financial statements. The US, Japanese, and many other governments require every listed firm to disclose its major clients whose sales account for more than 10\% of its total sales, at least in the period examined in this study \cite{FASF2024}. Therefore, based on published financial statements and, in addition, supplemental information from news articles and websites, FactSet identifies supply chains of listed and non-listed firms with their major partners. Table \ref{tab:sc_global} shows the number of all firms (for reference) and listed firms (our sample) in the FactSet data and their supply-chain links by year. The number of firms covered in the FactSet data increased rapidly over time, because the dataset has been constructed recently and thus does not cover firms in the distant past. 
In a later section, the recorded supply chain links at the time are used to examine news-related stock price changes along supply chain relationships.


The FactSet supply chain database does not cover the full universe of firms worldwide.
Instead, it identifies major customer–supplier relationships disclosed in financial statements and related sources, which are typically limited to economically significant partners. As a result, its coverage is intentionally limited, especially in earlier years.

FactSet uses its own identifier system, referred to as the FactSet Security Identifier. We convert these identifiers to RICs in order to link the FactSet data with news and stock price data.

\begin{table}[tb]
    \caption{Coverage of the FactSet supply chain database: number of firms and supply-chain links by year (global)}
    \label{tab:sc_global}
\footnotesize
    \begin{center}
    \begin{minipage}{.8\textwidth}
    \raggedright
    \begin{tabular}{l|ccccccc}
    \hline
        & 2003 & 2004 & 2005 & 2006 & 2007 & 2008 & 2009 \\ \hline
        All firms &&&&&& \\
        Number of firms & 9,283 & 9,292 & 9,804 & 10,058 & 10,342 & 10,197 & 10,802 \\
        Number of links & 38,594 & 40,749 & 45,172 & 38,866 & 34,894 & 33,136 & 28,640 \\
        Maximum indegree & 283 & 272 & 272 & 421 & 343 & 291 & 195 \\
        Maximum outdegree & 255 & 266 & 249 & 203 & 159 & 180 & 156 \\ \hline
        Listed &&&&&& \\
        Number of firms & 633 & 726 & 783 & 797 & 745 & 760 & 1,047 \\         
        Number of links & 1,413 & 1,685 & 1,914 & 1,723 & 1,386 & 1,364 & 1,687 \\
        Maximum indegree & 31 & 36 & 30 & 24 & 22 & 19 & 17 \\
        Maximum outdegree & 35 & 41 & 47 & 56 & 43 & 44 & 35 \\ \hline
    \end{tabular}
    \end{minipage}
    \vspace{10pt}
    \\
    \begin{minipage}{.8\textwidth}
    \raggedright
    \begin{tabular}{l|ccccccc}
    \hline
        & 2010 & 2011 & 2012 & 2013 & 2014 & 2015 & 2016 \\ \hline
        All firms &&&&&& \\
        Number of firms & 17,729 & 24,671 & 28,244 & 31,090 & 40,146 & 53,114 & 64,968 \\ 
        Number of links & 42,971 & 61,675 & 72,671 & 82,673 & 101,748 & 133,134 & 167,772 \\
        Maximum indegree & 463 & 600 & 577 & 496 & 423 & 586 & 562 \\
        Maximum outdegree & 162 & 201 & 206 & 198 & 256 & 294 & 378 \\ \hline
        Listed &&&&&& \\
        Number of firms & 2,153 & 3,842 & 4,798 & 5,779 & 8,418 & 10,729 & 12,728 \\         
        Number of links & 4,179 & 8,782 & 12,073 & 15,485 & 22,473 & 31,537 & 42,201 \\
        Maximum indegree & 27 & 70 & 115 & 127 & 161 & 225 & 277 \\
        Maximum outdegree & 40 & 54 & 79 & 79 & 75 & 109 & 140 \\ \hline
    \end{tabular}
    \end{minipage}
    \end{center}
\end{table}

To identify domestic supply chains within Japan, we employ data collected by Tokyo Shoko Research (TSR) for the period 2008--2016, particularly its Company Linkage Database that covers supply chains of more than one million firms in Japan. Supply chains are captured by annual firm-level surveys by TSR that request each firm to report up to 23 suppliers and client firms for each surveyed firm. Apparently, many firms are linked with more than 23 suppliers and clients, and thus their links cannot be fully identified by their own responses. However, these links can be mostly identified by responses by their suppliers and clients. Since those firms are relatively large, their partners report them as important suppliers or clients.
Therefore, although the numbers of suppliers and clients are limited to 23 in the reports, large firms can have many partners.
Accordingly, the maximum number of suppliers and clients is 12,729 and 12,016, respectively, in 2016, thus providing broad coverage of supply chains of Japanese firms. 

The TSR database provides exceptionally broad coverage of firm-level relationships, including both listed and unlisted firms.
In this study, however, we restrict our analysis to relationships involving listed firms and links between such firms, in order to ensure consistency with market-based information.
Listed firms in the TSR database are identified by corporate IDs and converted into RICs.
The number of firms and their supply-chain links and the corresponding number for listed firms from 2008 to 2016 are presented in Table \ref{tab:sc_japan}. 

\begin{table}[tb]
    \caption{Coverage of the Tokyo Shoko Research (TSR) supply chain database: number of firms and supply-chain links by year (Japan)}
    \label{tab:sc_japan}
\begin{center}

    \begin{minipage}{.7\textwidth}
\raggedright
    \footnotesize
    \begin{tabular}{l|ccccc}
    \hline
        & 2008 & 2009 & 2010 & 2011 & 2012  \\ \hline
        All firms &&&&& \\
        Number of firms & 1,018,570 & 1,061,051 & 1,107,760 & 1,135,200 & 1,156,205   \\ 
        Number of links & 4,443,677 & 4,644,977 & 4,810,967 & 4,925,307 & 4,995,612  \\
        Maximum indegree & 6,751 & 6,976 & 7,069 & 7,332 & 8,398  \\
        Maximum outdegree & 11,140 & 11,150 & 11,155 & 11,201 & 11,287  \\ \hline
        Listed  &&&&& \\
        Number of firms & 3,554 & 3,456 & 3,403 & 3,308 & 3,245 \\ 
        Number of links & 21,720 & 22,128 & 22,630 & 22,044 & 21,833 \\
        Maximum indegree & 201 & 189 & 180 & 162 & 151 \\
        Maximum outdegree & 305 & 277 & 259 & 243 & 225 \\ \hline
    \end{tabular}
       \end{minipage}

        \vspace{10pt}
    \begin{minipage}{.7\textwidth}
\raggedright
    \footnotesize
    \begin{tabular}{l|cccc}
    \hline
         & 2013 & 2014 & 2015 & 2016 \\ \hline
        All firms &&&& \\
        Number of firms & 1,182,729 & 1,194,034 & 1,200,162 & 1,247,939 \\ 
        Number of links & 5,090,119 & 5,157,181 & 5,209,604 & 5,491,417 \\
        Maximum indegree & 8,787 & 9,520 & 9,948 & 12,016 \\
        Maximum outdegree & 11,490 & 11,586 & 11,588 & 12,729 \\ \hline
        Listed  &&&& \\
        Number of firms & 3,215 & 3,204 & 3,215 & 3,205 \\ 
        Number of links & 21,696 & 21,439 & 20,968 & 20,445 \\
        Maximum indegree & 148 & 145 & 143 & 141 \\
        Maximum outdegree & 216 & 201 & 195 & 190 \\ \hline
    \end{tabular}
       \end{minipage}
\end{center}
\end{table}

\section{Methods}

\subsection*{Sentiment analysis using FinBERT}

A growing literature examines how news content itself is quantified and linked to stock market reactions. Early studies quantified news content using dictionary-based and word-frequency approaches to capture investor sentiment and market reactions. For example, \citet{tetlock2007giving} construct a sentiment measure based on word categories in financial news and show that media pessimism predicts short-run market returns and trading volume. Relatedly, \citet{mizuno2017novel} quantify the novelty and topicality of business news using word-level similarity measures and demonstrate that such text-based characteristics are strongly associated with subsequent stock market activity. In this field, neural networks are predominantly used. For example, hedge funds, such as Bridgewater Associates, utilize neural networks for predictions \cite{bloomberg2024bridgewater}. In academia, the transformer \cite{vaswani2017all} is widely used \cite{sonkiya2021stock,fazlija2022using}. In addition to the transformer, there is a long history of using neural networks to predict stock prices, such as long short-term memory \cite{jang2020business,ko2021lstm,dong2020belt,eapen2019novel,jin2020stock,li2018stock,selvin2017stock,fischer2018deep}, convolutional neural networks \cite{hoseinzade2019cnnpred}, or other deep learning methods \cite{maqsood2020local,patil2020stock,sim2019deep, oncharoen2018deep}. We use sentiment analysis to determine the positiveness and negativeness of each news article \cite{medhat2014sentiment,godbole2007large} and for this purpose, in particular, Bidirectional Encoder Representations from Transformers (BERT) \cite{devlin2018bert} have garnered the most acclaim \cite{sousa2019bert}. We take advantage of FinBERT \cite{araci2019finbert}, a fine-tuned model of BERT particularly for financial information. 

\begin{figure}[tb]
\centering
  \begin{minipage}[b]{0.45\columnwidth}
    \centering
    \includegraphics[width=\columnwidth]{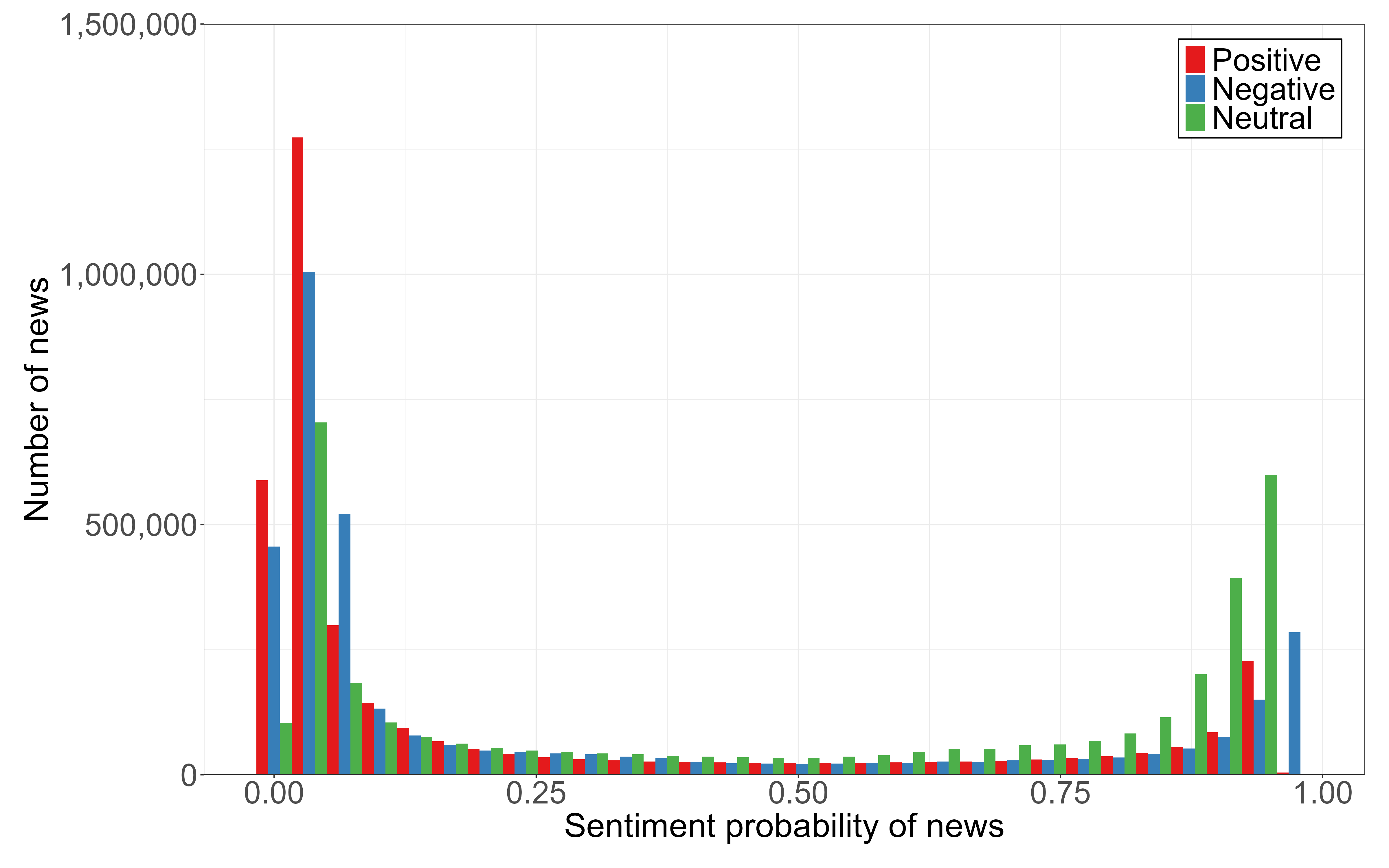}
    \caption*{(A) Firms across the world}
  \end{minipage}
  \hspace{0.01\columnwidth}
  \begin{minipage}[b]{0.45\columnwidth}
    \centering
    \includegraphics[width=\columnwidth]{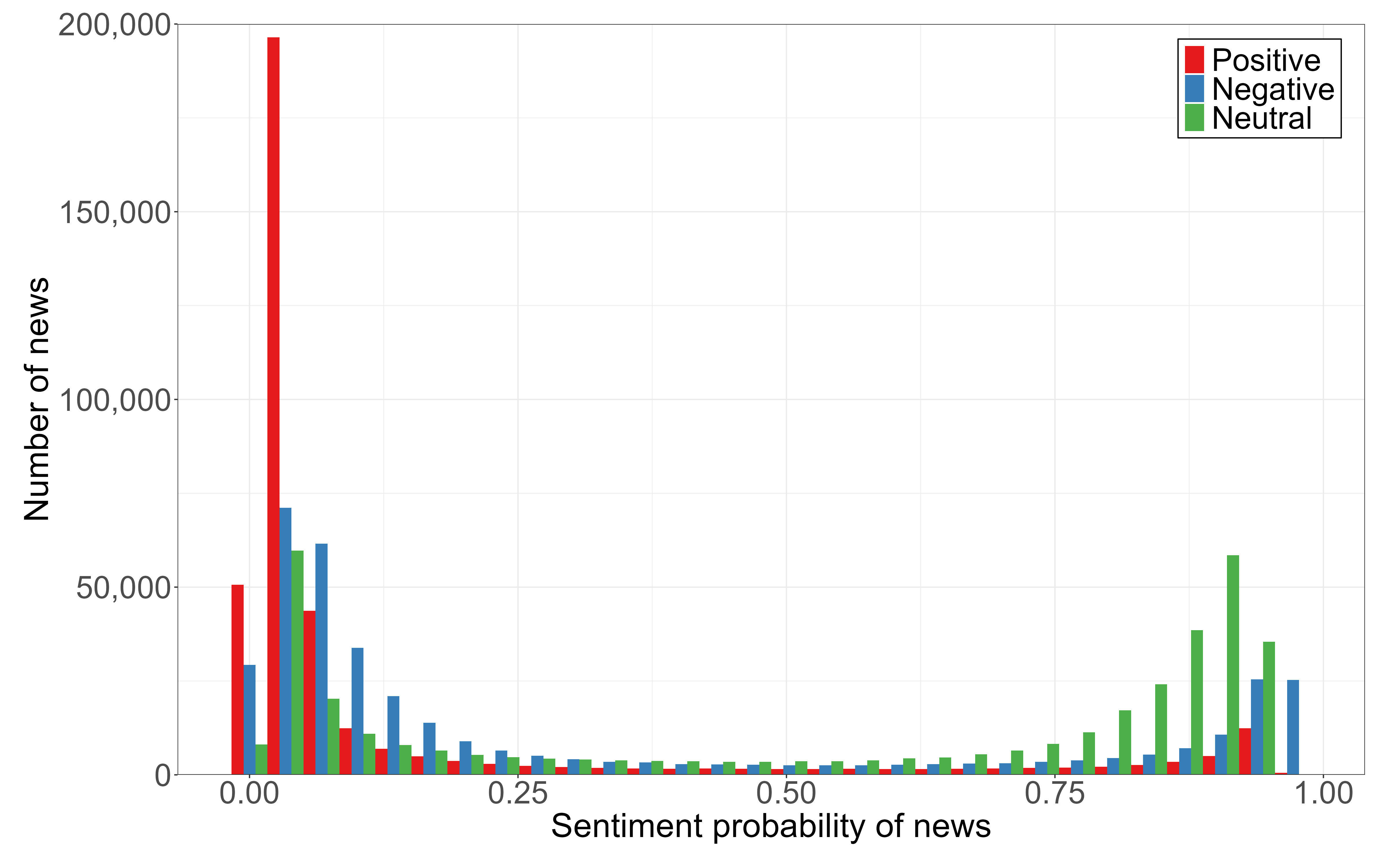}
    \caption*{(B) Japanese firms}
  \end{minipage}
  \caption{Distribution of the probabilities of positive, neutral, and negative sentiments in news articles generated by FinBERT.}
  \label{fig:sentiment}
\end{figure}

Our sentiment analysis to determine the degree of positive and negative sentiment of the information about the listed firms provided in each news article utilizes a natural language processing (NLP) model called FinBERT \cite{araci2019finbert}. FinBERT is a variant of the Bidirectional Encoder Representations from Transformers (BERT) developed by Devlin et al. (2018) \cite{devlin2018bert}. Specifically, FinBERT uses Reuters' TRC2-financial, which consists of 1.8 million news articles between 2008 and 2010 and fine-tunes BERT for financial sentiment classification using Financial Phrasebank, which consists of 4,845 English sentences taken from financial news in the LexisNexis database. From each news article, FinBERT generates softmax outputs for three labels, i.e., positive, neutral, and negative, which indicate the weights of the three sentiments about the firms mentioned in the article. Examples are shown in Table \ref{tab:news}. Before conducting the sentiment analysis using FinBERT, we clean news articles by deleting URLs, line breaks, International Securities Identification Numbers, and fixed phrases at the end. 
The distributions of the probabilities of positive, neutral, and negative sentiments generated by FinBERT from news articles about firms across the world and Japanese firms are provided in Panels (A) and (B) of Figure \ref{fig:sentiment}, respectively. Both panels indicate that the probability of neutral sentiment is high for many news articles followed by that of negative sentiment while news articles with a large probability of positive sentiment are fewer.   

From the three sentiment probabilities generated by FinBERT (positive, neutral, and negative),
we construct sentiment measures by excluding the neutral component and normalizing the remaining
positive and negative probabilities.
Specifically, the positive sentiment index is defined as
\begin{equation}
    \text{Positive}_{int} = \frac{p^{+}_{int}}{p^{+}_{int} + p^{-}_{int}},
\end{equation}
and the negative sentiment index as
\begin{equation}
    \text{Negative}_{int} = \frac{p^{-}_{int}}{p^{+}_{int} + p^{-}_{int}},
\end{equation}
where $p^{+}_{int}$ and $p^{-}_{int}$ denote the probabilities of positive
and negative sentiment, respectively, for news article $n$ mentioning firm $i$ on day $t$.
This normalization focuses the analysis on the relative balance between positive and negative
information and avoids ambiguity associated with neutral sentiment.

\begin{table}[tb]
    \caption{Examples of news articles and sentiments.}
    \label{tab:news}
\footnotesize
    \centering
    \begin{tabular}{|p{10cm}|ccc|}
    \hline
     & \multicolumn{3}{c|}{Sentiment} \\
     \cline{2-4}
    News article & \multicolumn{1}{l}{Positive} & \multicolumn{1}{l}{Neutral} & \multicolumn{1}{l|}{Negative} \\
    \hline
    (Sep/11/2008) Advanced Medical Solutions Group Plc on Thursday said its silver anti-microbial wound gel had been approved by the U.S. Food \& Drug Administration (FDA), sending its shares up 5 percent. ... [Source: \cite{reuters2008advanced}]
    & 0.93 & 0.01 & 0.06 \\
    \hline
    (Sep/11/2008) U.S. Interior Department employees who oversaw oil drilling on federal lands had sex and used illegal drugs with workers at energy companies ... [Source: \cite{reuters2008usgov}]
    & 0.02 & 0.21 & 0.76 \\
    \hline
    (Sep/11/2008) Amazon.com, the largest global online retailer, plans to start selling U.S.-produced wine on its website within the United States by late September or October ... [Source: \cite{reuters2008amazon}]
    & 0.14 & 0.84 & 0.01 \\
    \hline
    \end{tabular}
\end{table}

\subsection{Regression analysis}

\subsubsection*{Own-firm associations}

To examine the association between positive and negative sentiment in news articles about particular firms and their own stock prices, we apply the positive and negative sentiment indexes constructed above in regression analysis. Specifically, we focus on firms mentioned by news articles and examine whether the stock price change rate in a time window is associated with the level of positiveness or negativeness of the article in the pre- and post-disclosure periods.

\begin{figure}[tb]
\centering
\includegraphics[width=.7\linewidth]{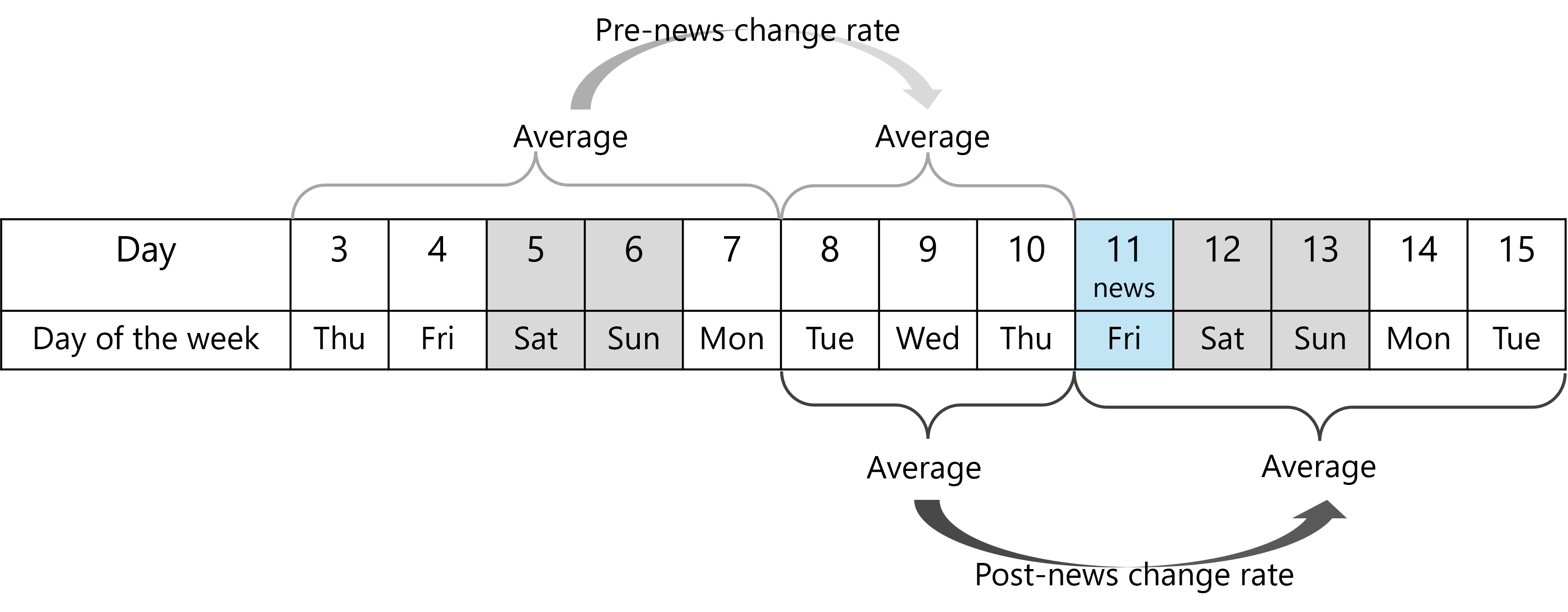}
\caption{Definition of pre- and post-news change rate of stock prices. An example when a news article is disclosed on Friday, the 11th of June and the time window is 3 days.}
\label{fig:def_change}
\end{figure}

Suppose that firm $i$ listed in stock market $m$ is mentioned by news article $n$ on day $t$. Setting the time window at $w$, such as 1, 5, and 30 days, we define the average daily percentage change of the stock price of firm $i$ before and after news article $n$ was disclosed on day $t$ in time window $w$, or $(\%\Delta P/P)^{\mathit{pre}}_{intw}$ and $(\%\Delta P/P)^{\mathit{post}}_{intw}$, respectively, by 
\begin{align}
(\%\Delta P/P)^{\mathit{pre}}_{intw} &=
(\ln \bar{P_i}[t-w,t-1] - \ln \bar{P_i}[t-2w,t-w-1])/w \times 100 &&\text{for the pre-news change}, 
\label{eq:prate_pre}\\ 
(\%\Delta P/P)^{\mathit{post}}_{intw} &=
(\ln \bar{P_i}[t, t+w-1] - \ln \bar{P_i}[t-w,t-1])/w \times 100 &&\text{for the post-news change},
\label{eq:prate_post}
\end{align}
where $\bar{P_i}[t_1,t_2]$ is the average stock price of firm $i$ between day $t_1$ and $t_2$. For example, $\bar{P_i}[t-w,t-1]$ is the average stock price of firm $i$ for $w$ days between day $t-w$ and $t-1$. Therefore, the left-hand side of equations (\ref{eq:prate_pre}) and (\ref{eq:prate_post}) indicates the daily percentage change of the average stock price in time window $w$ in the pre- and post-news period, respectively. We take the average of stock prices over the time window to reduce their fluctuations. We skip days when stock prices are unavailable, such as Saturdays, Sundays, and holidays.
Accordingly, for example, as shown in Figure \ref{fig:def_change}, when news article $n$ about firm $i$ is disclosed on Friday, the 11th of June and the time window is set at 3 days, the pre-news change rate is defined as the change rate from the average stock price on June 3rd (Thursday), 4th (Friday), and 7th (Monday), skipping 5th (Saturday) and 6th (Sunday), to the average from the 8th (Tuesday) to 10th (Thursday). Similarly, the post-news change is defined as the change from the average of June 8--10 to that of June 11th, 14th (Monday), and 15th (Tuesday).  

Accordingly, the unit of observation in our analysis is a firm–news-article pair. For each firm mentioned in a news article, we construct two observations corresponding to the pre- and post-disclosure periods. As a result, the total number of observations equals twice the cumulative number of firm mentions in news articles over the sample period.

Using the pre- and post-news observations, we estimate the following estimation equation:
\begin{equation}
\begin{split}
     (\%\Delta P/P)^{T}_{\mathit{intw}}  
        = \beta^{pre}_w (\mathit{PRE}_{intw} \times \mathit{NEWS}_{int}) + \beta^{post}_w (\mathit{POST}_{intw} \times \mathit{NEWS}_{int}) 
        + \beta_X X_{mntw} + \mu_s + \epsilon_{intw},
\end{split}
\label{eq:reg1}
\end{equation}
where T is either \textit{pre} or \textit{post}, and $\mathit{PRE}{intw}$ and $\mathit{POST}{intw}$ are indicator variables that take a value of one if $(\%\Delta P/P)^T_{intw}$ represents the percentage change in the stock price of firm $i$ over the length of the time window $w$ before or after news article $n$ is disclosed on day $t$, respectively.
$\mathit{NEWS}_{int}$ represents the level of positive or negative sentiment in news article $n$
generated by FinBERT and corresponds to the sentiment indexes
$\text{Positive}_{int}$ and $\text{Negative}_{int}$ defined in the Methods section.
$X_{mntw}$ is the percentage change of the Refinitiv index for each stock market $m$ that controls for the overall time trend of stock prices in the market, and $\mu_s$ is sector dummies. Sectors are defined by the North American Industry Classification System at the sub-sector level for the sample of firms across the world and by the Japan Standard Industrial Classification at the two-digit level for the sample of Japanese firms. 

Accordingly, $\beta^{\mathit{post}}_w$ indicates the average association between the positiveness or negativeness of a news article and the average change rate of the stock price of the firm mentioned in the article after it is disclosed in the time window of $w$ days, controlling for the overall trend in the market and sector-specific unobservable factors. By contrast, $\beta^{\mathit{pre}}_w$ captures any prior association between the news article and the change rate of the stock price before disclosure. This term may reflect a combination of factors that are not separately identified in this analysis, including selection into news coverage and pre-existing price dynamics. In addition, $\beta^{\mathit{pre}}_w$ may include the endogeneity bias reflecting the possibility that the growth rate of the stock price of a firm that is positively mentioned by any news article is intrinsically higher than that of a firm negatively mentioned. Accordingly, the difference between the post- and pre-news coefficients, $\beta^{\mathit{post}}_w - \beta^{\mathit{pre}}_w$, summarizes changes in stock price growth around the time of news disclosure \citep{isaksson2018china}, relative to the pre-disclosure period. This difference compares post-disclosure price movements with those observed before disclosure,
where pre-news movements may reflect the incorporation of information prior to public disclosure.

We particularly estimate equation (\ref{eq:reg1}) for $w = $ 1, 2, 3, 4, 5, 30, 180, and 365 to examine both the short- and long-run associations between positive news articles and stock prices. We hypothesize that a positive (negative) news article is associated with higher (lower) stock price changes in the short-run, with weaker associations in the longer run. The analysis is conducted on the sub-sample of firms mentioned in news articles, as firms that are mentioned and those that are not mentioned are likely to differ systematically in ways that are difficult to control for.

\subsubsection*{Associations along supply chains}

We further examine the association between positive or negative sentiment about firms in news articles and stock price changes of their suppliers and clients, using analogous frameworks. Specifically, when we focus on upstream associations between news articles about firms and their suppliers, the variable $\mathit{NEWS}_{int}$ in equation (\ref{eq:reg1}) is replaced with the positive or negative sentiment indixes from news articles that mention any of firm $i$’s client firms. Alternatively, when we examine the downstream associations between news articles about firms and their clients, $\mathit{NEWS}_{int}$ is replaced with the positive or negative indixes of any of the suppliers of firm $i$. When a firm mentioned by a news article has more than one supplier or client, as shown in Tables \ref{tab:sc_global} and \ref{tab:sc_japan}, we include all the suppliers and clients in our sample. 

We hypothesize that news sentiment about a firm is associated with stock price changes of its suppliers and clients. These associations may reflect, for example, investors’ expectations about changes in demand faced by suppliers or about the performance of downstream firms, although such mechanisms are not directly tested in this study. As in the case of own-firm associations, these associations are expected to be stronger over shorter horizons and weaker over longer horizons.

\section{Results}

\subsection*{Own-firm stock price associations for listed firms across the world}

\begin{figure}[tb]
\begin{minipage}{0.48\textwidth}
    \centering
    \includegraphics[width=\textwidth]{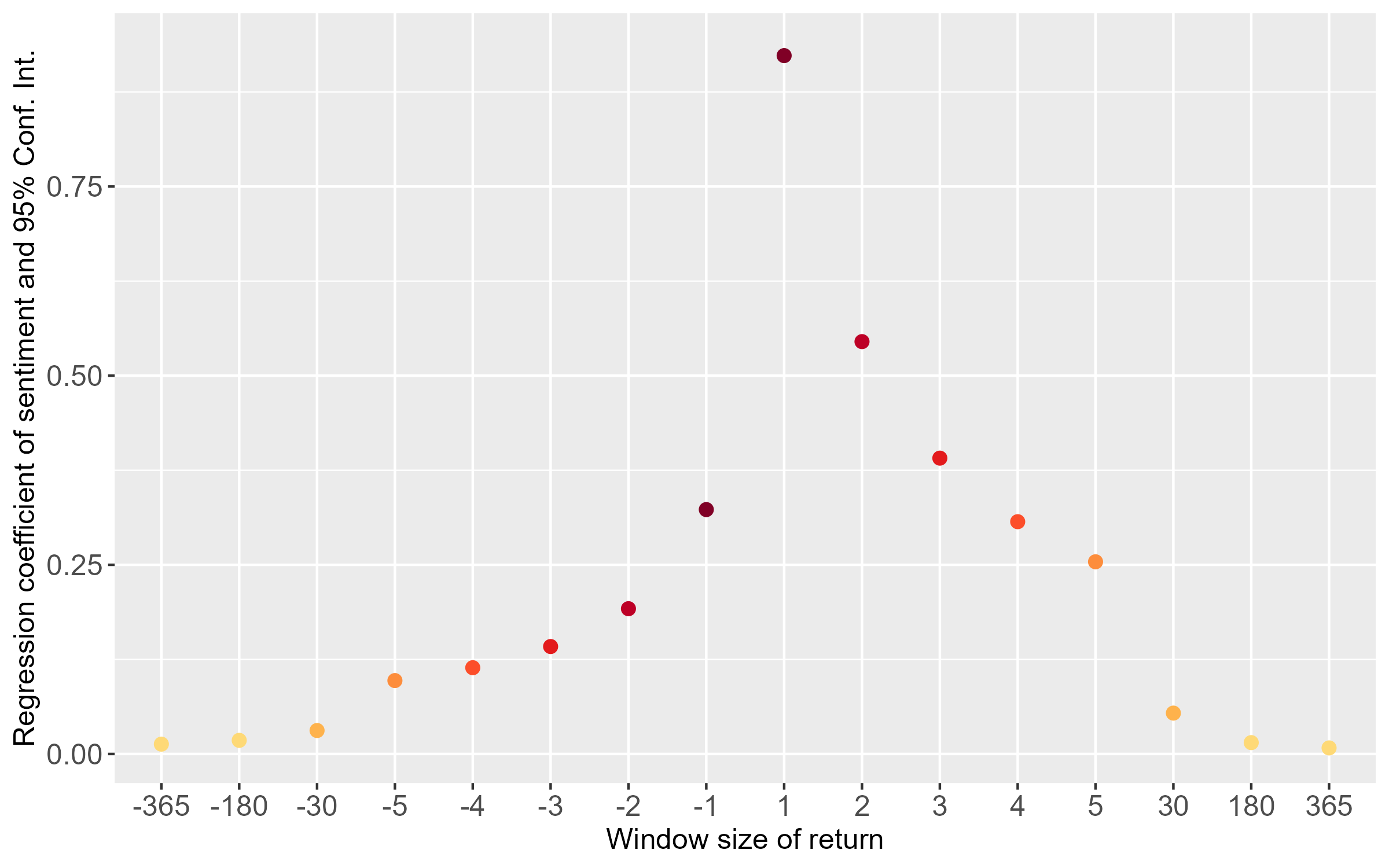} 
    \caption*{(A) Coefficient estimates of positive sentiment} 
\end{minipage}
\hfill
\begin{minipage}{0.48\textwidth}
    \centering
    \includegraphics[width=\textwidth]{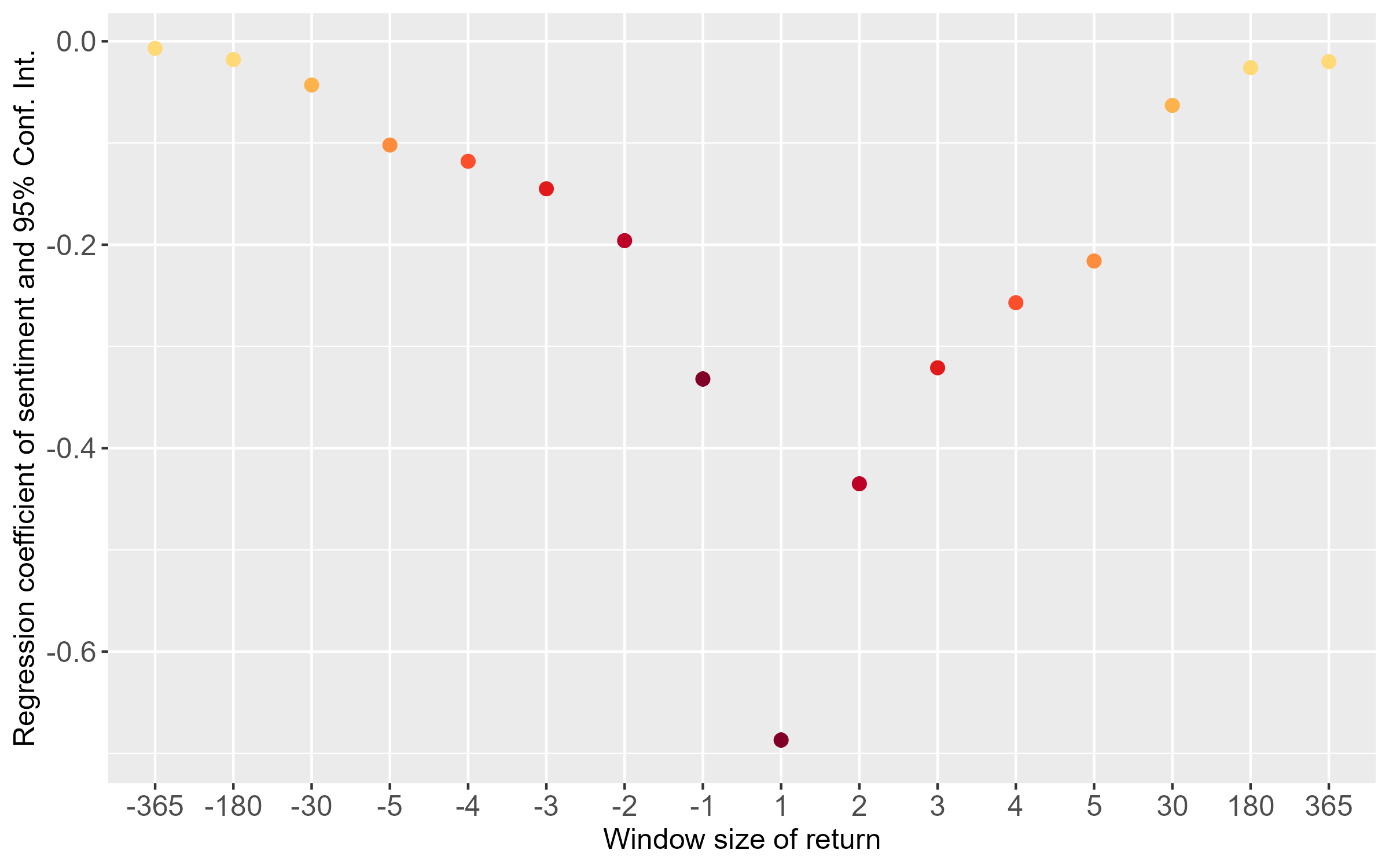}
    \caption*{(B) Coefficient estimates of negative sentiment} 
\end{minipage}
\caption{Coefficient estimates of the positive (panel [A]) and negative (panel [B]) sentiment indexes of news articles about \underline{firms across the world} on percentage changes of their \underline{own} stock prices for different time windows. Panel (A) is a graphical presentation of Table \ref{tab:reg_global_own}. When the value of the horizontal axis is negative and $-w$, the dot above the value indicates the pre-news coefficient estimate for time window $w$ ($\beta^{\mathit{pre}}_w$). When it is positive and $w$, the dot indicates the post-news coefficient estimate for time window $w$ ($\beta^{\mathit{post}}_w$). The color of the dots for time windows $w$ and $-w$ is set to be the same so that the post- and pre-news coefficients can be easily compared. The confidence interval at the 5\% level associated with each point estimate (dot) is shown by a vertical segment but invisible because the confidence intervals are negligible compared with the point estimate.}
\label{fig:reg_global_own}
\end{figure}

We start by examining how the positiveness of news articles about firms is associated with percentage changes in their stock prices before and after the disclosure of news, applying equation (\ref{eq:reg1}) to the sample of publicly listed firms across the world mentioned by any news article. Results are presented in Table \ref{tab:reg_global_own} and Figure \ref{fig:reg_global_own}. Then, we show the results for the negativeness.

\begin{table}[tb]
   \caption{Coefficient estimates for the positiveness index of news articles about \underline{firms across the world} on percentage changes of their \underline{own} stock prices before and after the disclosure of news for different time windows ($\beta^{\mathit{pre}}_w$ and $\beta^{\mathit{post}}_w$ in equation [\ref{eq:reg1}], respectively). Standard errors are in parentheses. ***, **, and * indicate statistical significance at the 1, 5, and 10\% level, respectively. The second row from the bottom shows $p$ values from $t$ tests for $\beta^{\mathit{post}}_w - \beta^{\mathit{pre}}_w =0$.}
   \label{tab:reg_global_own}
\vspace{-5ex}
\begin{center}
\scalebox{0.6}{
\begin{tabular}{lcccccccc}
   \tabularnewline \midrule \midrule
    Dependent Variable: & \multicolumn{7}{c}{Daily percentage change of the stock price of the firm mentioned in each news article in the pre- or post-news period}\\
   \midrule
   Model: & (1) & (2) & (3) & (4) & (5) & (6) & (7) & (8)\\  
   Time window (days) [$w$] & 1 & 2 & 3 & 4 & 5 & 30 & 180 & 365\\   
   \midrule
   $\beta^{\mathit{pre}}_w$ ($\mathit{PRE}_{intw} \times \mathit{NEWS}_{int}$) & $3.23\times 10^{-1}$$^{***}$  & $1.92\times 10^{-1}$$^{***}$  & $1.42\times 10^{-1}$$^{***}$  & $1.14\times 10^{-1}$$^{***}$  & $9.72\times 10^{-2}$$^{***}$  & $3.13\times 10^{-2}$$^{***}$  & $1.78\times 10^{-2}$$^{***}$  & $1.33\times 10^{-2}$$^{***}$\\    
   & ($4.82\times 10^{-3}$) & ($2.74\times 10^{-3}$) & ($2.05\times 10^{-3}$) & ($1.67\times 10^{-3}$) & ($1.44\times 10^{-3}$) & ($4.77\times 10^{-4}$) & ($1.77\times 10^{-4}$) & ($1.23\times 10^{-4}$)\\    
   $\beta^{\mathit{post}}_w$ ($\mathit{POST}_{intw} \times \mathit{NEWS}_{int}$)   & $9.23\times 10^{-1}$$^{***}$  & $5.45\times 10^{-1}$$^{***}$  & $3.91\times 10^{-1}$$^{***}$  & $3.07\times 10^{-1}$$^{***}$  & $2.54\times 10^{-1}$$^{***}$  & $5.41\times 10^{-2}$$^{***}$  & $1.48\times 10^{-2}$$^{***}$  & $7.92\times 10^{-3}$$^{***}$\\    
   & ($4.89\times 10^{-3}$) & ($2.75\times 10^{-3}$) & ($2.05\times 10^{-3}$) & ($1.68\times 10^{-3}$) & ($1.44\times 10^{-3}$) & ($4.77\times 10^{-4}$) & ($1.76\times 10^{-4}$) & ($1.21\times 10^{-4}$)\\    
   \midrule
   Market-specific trend & Yes & Yes & Yes & Yes & Yes & Yes & Yes & Yes \\  
   Sector fixed effects & Yes & Yes & Yes & Yes & Yes & Yes & Yes & Yes\\  
   \midrule
   $\beta^{\mathit{post}}_w - \beta^{\mathit{pre}}_w$ & $0.600^{***}$ & $0.353^{***}$ & $0.249^{***}$ & $0.193^{***}$ & $0.157^{***}$ & $0.023^{***}$ & $-0.003^{***}$ & $-0.005$\\
   $p$ value & 0.00 & 0.00 & 0.00 & 0.00 & 0.00 & 0.00 & 0.00 & 0.00 \\
   \midrule
   Number of observations   & 9,409,978                     & 9,543,586                     & 9,562,439                     & 9,565,054                     & 9,562,897                     & 9,460,176                     & 8,986,209                     & 8,546,395\\  
   \midrule \midrule
\end{tabular}
}
\end{center}
\end{table}

In column (1) of Table \ref{tab:reg_global_own}, the estimated $\beta^{\mathit{pre}}_1$ is 0.323 and highly significant, indicating that a higher value of the positive sentiment index in a news article about a firm is associated with the percentage change in the firm's stock price from two days before to one day before the release of the news. Since the positiveness measure, $\mathit{NEWS}$ in equation (\ref{eq:reg1}), ranges from 0 to 1, the value of the coefficient means that the change rate of the stock price of a firm mentioned in a news article most positively ($\mathit{NEWS} = 1$) is 0.323\% points higher than that of a firm mentioned with no positive sentiment ($\mathit{NEWS} = 0$) before the disclosure of the news. There are two possibilities for this positive coefficient. First, positive information, i.e., actual positive incident is incorporated into stock prices prior to  disclosure of news. Second, stock prices of firms positively mentioned by news articles exhibits higher intrinsic growth than those of firms with no positive mention, leading to endogeneity biases. In the following analysis, we do not attempt to disentangle these two channels. Instead, we focus on the difference between post- and pre-news coefficient estimates, which summarizes the additional change in stock prices around public news disclosure relative to pre-existing price movements.

Column (1) of Table \ref{tab:reg_global_own} further shows that the estimated $\beta^{\mathit{post}}_1$ is 0.923, indicating that, in the post-news period, firms mentioned with positive news sentiment experience stock price growth that is higher by 0.923 percentage points than firms mentioned with no positive sentiment. The second and third rows from the bottom show $\beta^{\mathit{post}}_w - \beta^{\mathit{pre}}_w$ and the $p$ value from a $t$ test for the null hypothesis that the difference is 0. In column (1), the $p$ value is close to 0, indicating that the difference is statistically different from zero. This result suggests that although stock prices of firms mentioned by news positively grow faster than those of firms mentioned with no positive sentiment even before the disclosure of news, stock prices of firms with positive news sentiment are higher by an additional 0.6\% after its disclosure. 

These results in column (1) for the time window of 1 day generally hold in columns (2)--(6) for time windows of 2--5 and 30 days, while the post- and pre-news coefficient estimates and the difference between the two decrease as the time window becomes wider. By contrast, the difference between the post- and pre-news coefficient estimates is negative when the time window is 180 and 365 days, or in the long run, although the difference is quite small in size. This result implies that the intrinsic difference between firms mentioned positively and those with no positive mention in news articles is negligible.

Panel (A) of Figure \ref{fig:reg_global_own} graphically demonstrates the argument above. The left half of the figure where the value on the horizontal axis is negative shows the point estimate of the pre-news coefficient for different time windows, whereas the right half shows the post-news coefficient. Although the confidence intervals are added in the figure, they are too small to be recognized. The estimated coefficients show an inverted-U shape where the coefficients are close to 0 on the left and right edges. This pattern indicates the following. The change rate of the stock price of firms mentioned with positive news sentiment is not intrinsically different from that of firms mentioned with no positive news sentiment long before the disclosure of the news. However, the change rate of the stock price of positively mentioned firms began to rise compared to firms with no positive mention 30 days before the disclosure, peaking one day after the disclosure. Then, the difference in the change rates of stock prices between positively mentioned firms and those with no positive mention gradually decreases over time, converging to nearly zero 180 days after the disclosure.

The pattern of coefficients for negative sentiment shown in Panel (B) of Figure \ref{fig:reg_global_own} is qualitatively consistent with the pattern of positive sentiment shown in Panel (A). The corresponding detailed regression results are reported in the Supplementary Information (SI Table 3). The coefficients are negligible 180 days or more before its disclosure but become negative and significant 30 days or fewer before the disclosure. After disclosure, the magnitude of the coefficients increase further. 

\subsection*{Associations for suppliers and clients along global supply chains}

Next, we examine whether news sentiment about a firm is associated with stock prices of the firm's suppliers and clients along global supply chains by estimating equation (\ref{eq:reg1}) where the positiveness or negativeness index of news article $n$ about firm $i$ on day $t$, $\mathit{NEWS}_{int}$, is replaced with the index for any supplier or client of firm $i$. 

\begin{figure}[tb]
\begin{minipage}{0.48\textwidth}
    \centering
    \includegraphics[width=\textwidth]{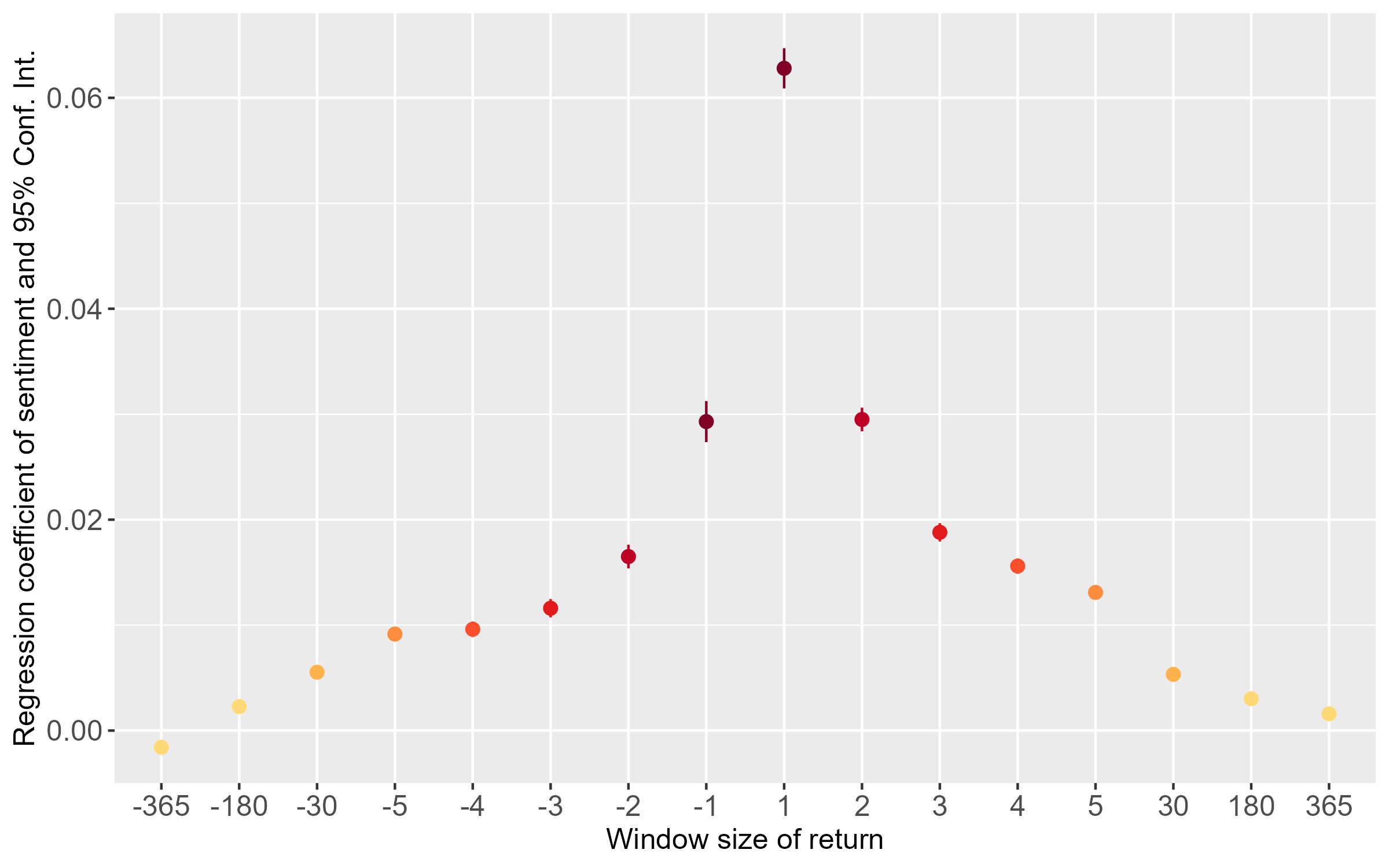} 
    \caption*{(A) Coefficient estimates of positive sentiment} 
\end{minipage}
\hfill
\begin{minipage}{0.48\textwidth}
    \centering
    \includegraphics[width=\textwidth]{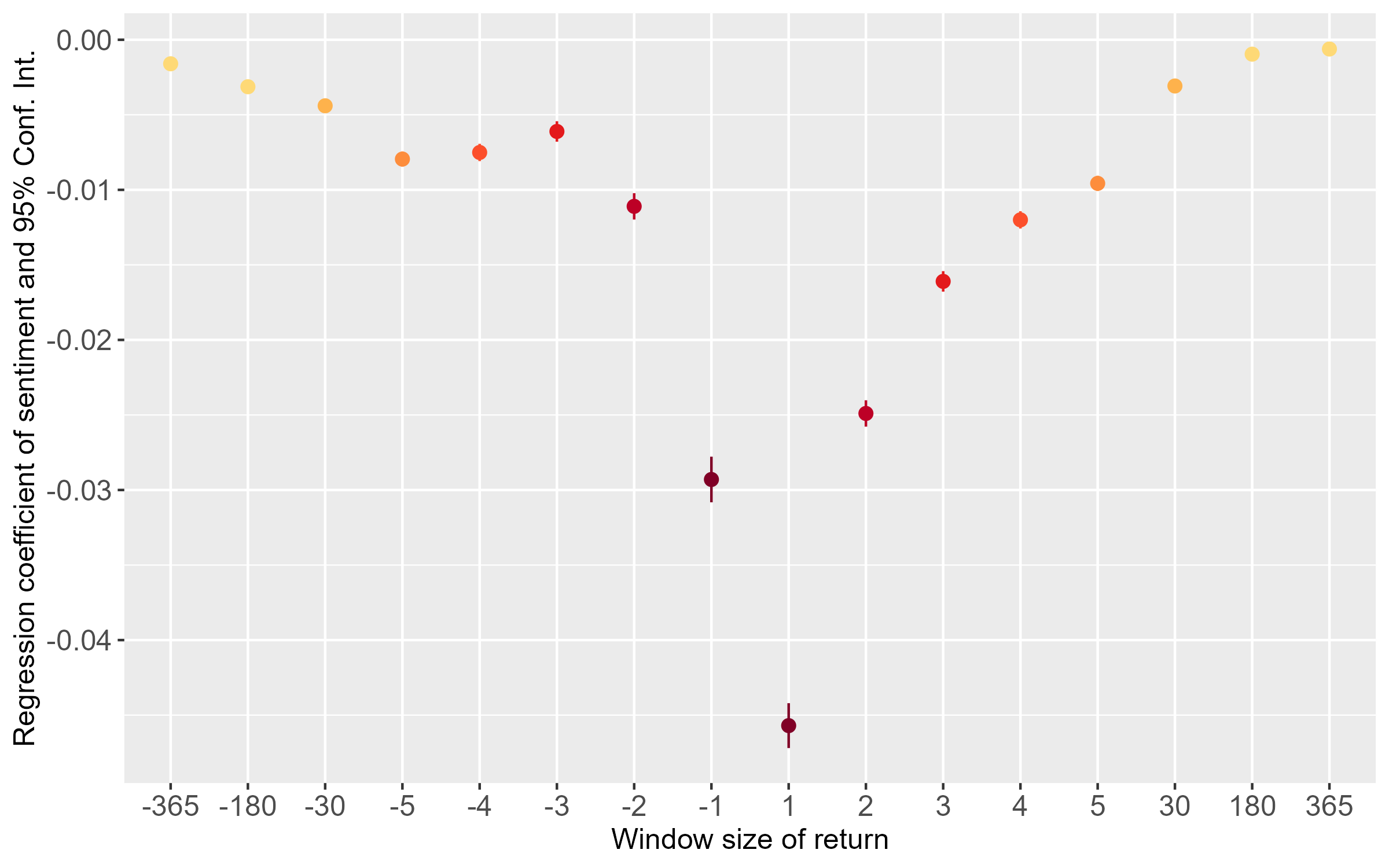}
    \caption*{(B) Coefficient estimates of negative sentiment} 
\end{minipage}
\caption{Coefficient estimates of the positive (panel (A)) and negative (panel (B) sentiment indexes of news articles about \underline{firms across the world} on percentage changes of stock prices of their \underline{suppliers} for different time windows. When the value of the horizontal axis is negative and $-w$, the dot above the value indicates the point estimate of the pre-news effect for time window $w$ ($\beta^{\mathit{pre}}_w$). When it is positive and $w$, the dot indicates the post-news effect for time window $w$ ($\beta^{\mathit{post}}_w$). The color of the dots for time windows $w$ and $-w$ is set to be the same so that the post- and pre-news coefficients can be easily compared. The confidence interval at the 5\% level associated with each point estimate (dot) is shown by a vertical segment.}
\label{fig:reg_global_supplier}
\end{figure}

Panels (A) and (B) of Figure \ref{fig:reg_global_supplier} present coefficient estimates for positive and negative news sentiment about firms and the associated stock price changes of their suppliers. Detailed regression results are reported in SI Table 4 for positive sentiment and SI Table 5 for negative sentiment, respectively. Table \ref{tab:reg_global_supplier} reports the difference between post- and pre-news coefficient estimates for positive sentiment, with corresponding results for negative sentiment reported in SI Table 1. 

For suppliers, both pre- and post-news coefficient estimates are positive (except for the -365-day window) and statistically significant for positive news sentiment, with a larger post–pre difference for shorter time windows. The corresponding coefficient estimates for negative sentiment exhibit the opposite sign and similar magnitudes.

\begin{table}[tb] 
\caption{Difference between the post- and pre-news coefficient estimates for  positive news sentiment about \underline{firms across the world} on their \underline{suppliers}' stock prices. $P$ values are those from $t$ tests for the null hypothesis that the difference is 0. ***, **, and * indicate statistical significance at the 1, 5, and 10\% level, respectively.}
\label{tab:reg_global_supplier}
\centering
\scalebox{0.75}{
\begin{tabular}{lcccccccc}
\toprule
Time window (days)& 1 & 2 & 3 & 4 & 5 & 30 & 180 & 365 \\
\midrule
$\beta^{post}_w - \beta^{pre}_w$ & $0.033^{***}$ & $0.013^{***}$ & $0.007^{***}$ & $0.006$ & $0.004^{***}$ & $-0.00$ & $0.001^{**}$ & $0.003^{***}$ \\
$p$ value & 0.00 & 0.00 & 0.00 & 0.00 & 0.00 & 0.27 & 0.00 & 0.00 \\
\bottomrule
\end{tabular}
}
\end{table}

There are two notable differences between the own-firm coefficients and those for suppliers. First, the post- and pre-news coefficients for suppliers and their difference are substantially smaller than the corresponding own-firm coefficients and their difference. For example, the pre-news coefficient for suppliers for the time window of 1 day is 0.0293 (SI Table 4), while the corresponding coefficient for own firms is 0.323 (Table \ref{tab:reg_global_own}), more than 10 times larger. Similarly, the difference between the post- and pre-news coefficients, $\beta^{post}_w - \beta^{pre}_w$, is 0.033 for suppliers when the time window is one day (Table \ref{tab:reg_global_supplier}), while the corresponding difference for own firms is 0.600 (Table \ref{tab:reg_global_own}). Quantitatively, although positive news articles about firms are associated with stock prices of their suppliers, the corresponding coefficients for suppliers are less than 10\% of those for own firms. Second, the difference between the post- and pre-news coefficients on suppliers is 0.033 when the time window is 1 day but becomes 0.004 when it is 5 days, indicating that the post-pre difference in coefficients for suppliers declines by 88\% in 5 days. By contrast, the coefficients for own firm decline from 0.600 to 0.157 by 74\%. This comparison shows that the difference between the post- and pre-news coefficients for suppliers diminishes more quickly after disclosure than the corresponding difference for firms’ own stock prices. 

As in the analysis for firms’ own stock prices, negative sentiment shows a pattern similar to positive sentiment for suppliers (SI Table 5). Both the post- and pre-news coefficients for suppliers, and their difference, are substantially smaller than the corresponding coefficients for firms’ own stock prices, and the post–pre difference for suppliers diminishes more quickly after disclosure.

\begin{figure}[tb]
\begin{minipage}{0.48\textwidth}
    \centering
    \includegraphics[width=\textwidth]{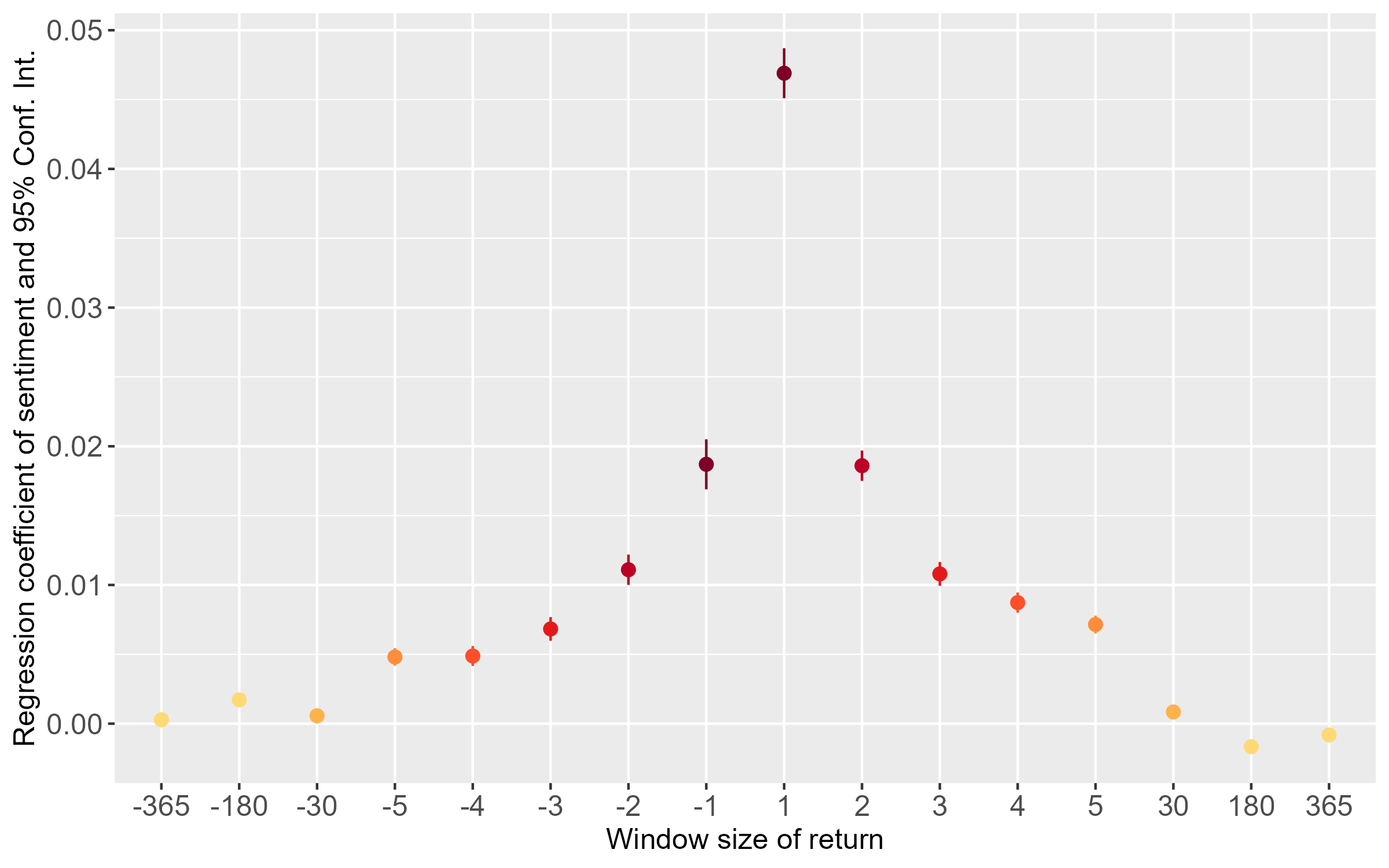} 
    \caption*{(A) Coefficient estimates of positive sentiment} 
\end{minipage}
\hfill
\begin{minipage}{0.48\textwidth}
    \centering
    \includegraphics[width=\textwidth]{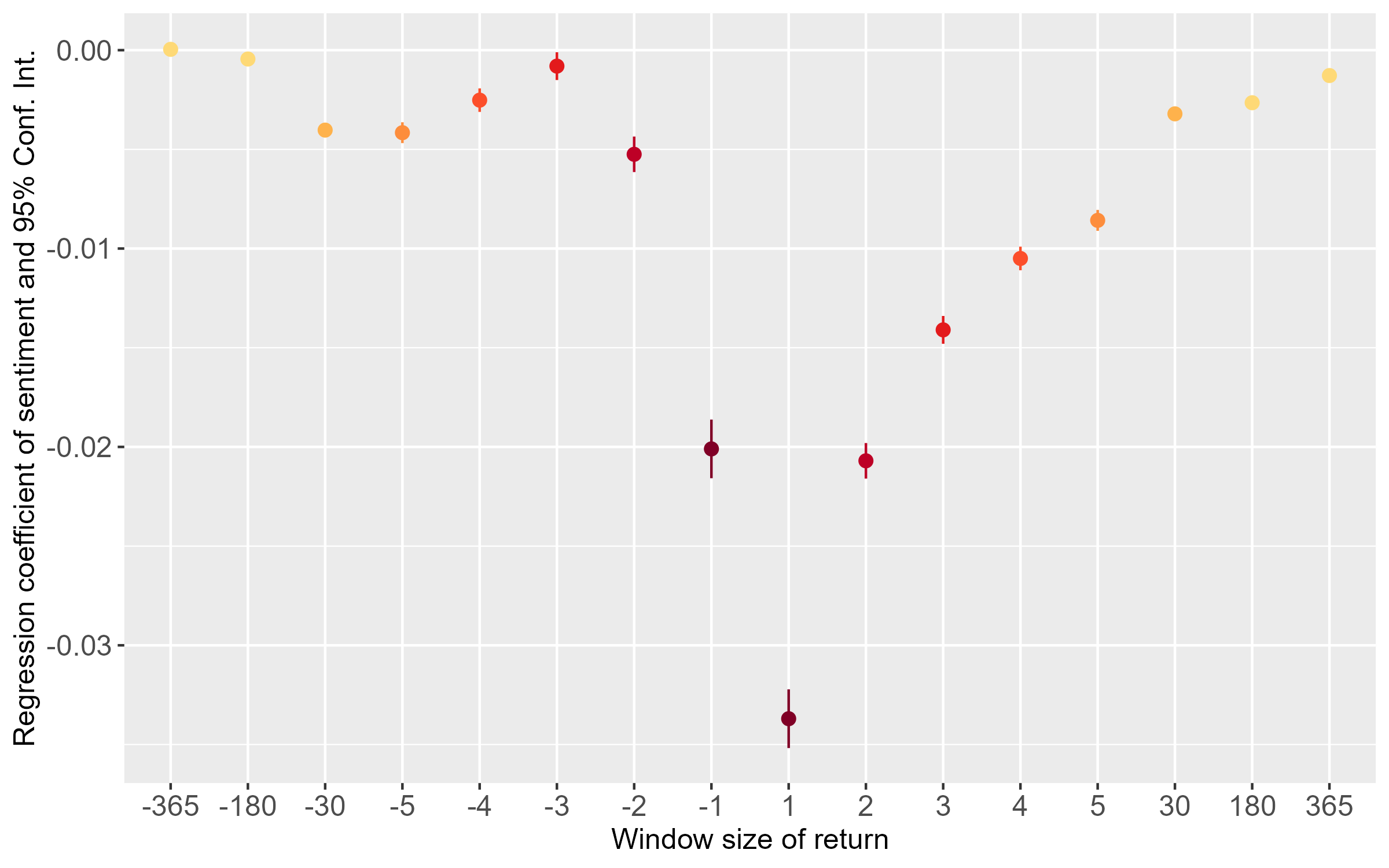}
    \caption*{(B) Coefficient estimates of negative sentiment} 
\end{minipage}
\caption{Coefficient estimates for the positive (panel (A)) and negative (panel (B)) sentiment indexes of news articles about \underline{firms across the world} on percentage changes of stock prices of their \underline{clients} for different time windows. When the value of the horizontal axis is negative and $-w$, the dot above the value indicates the point estimate of the pre-news coefficient for time window $w$ ($\beta^{\mathit{pre}}_w$). When it is positive and $w$, the dot indicates the post-news coefficient for time window $w$ ($\beta^{\mathit{post}}_w$). The color of the dots for time windows $w$ and $-w$ is set to be the same so that the post- and pre-news coefficients can be easily compared. The confidence interval at the 5\% level associated with each point estimate (dot) is shown by a vertical segment.}
\label{fig:reg_global_client}
\end{figure}

Similarly, we examine coefficient estimates for positive and negative news sentiment in relation to stock prices of client firms and present the results in Panels (A) and (B) of Figure \ref{fig:reg_global_client}, respectively. Detailed regression results for clients are reported in SI Table 6 for positive sentiment and in SI Table 7 for negative sentiment. We also show the difference between the post- and pre-news coefficient estimates for positive (Table \ref{tab:reg_global_client}) and negative (SI Table 2). These results exhibit the same qualitative pattern as those for suppliers shown in Figure \ref{fig:reg_global_supplier} and Table \ref{tab:reg_global_supplier}. Specifically, we find positive (negative) and significant pre- and post-news coefficients for positive (negative) news sentiment, as well as a positive (negative) and significant post–pre difference. However, the coefficients for clients are substantially smaller and decline more rapidly than those for own firms.

\begin{table}[tb] 
\caption{Difference between the post- and pre-news coefficient estimates of positive news sentiment about \underline{firms across the world} on their \underline{clients}' stock prices. $P$ values are those from $t$ tests for the null hypothesis that the difference is 0. Standard errors are in parentheses. ***, **, and * indicate statistical significance at the 1, 5, and 10\% level, respectively.}
\label{tab:reg_global_client}
\centering
\scalebox{0.75}{
\begin{tabular}{lcccccccc}
\toprule
Time window (days)& 1 & 2 & 3 & 4 & 5 & 30 & 180 & 365 \\
\midrule
$\beta^{post}_w - \beta^{pre}_w$ & $0.028^{***}$ & $0.007^{***}$ & $0.004^{***}$ & $0.004^{***}$ & $0.002^{***}$ & $0.000$ & $-0.003^{**}$ & $-0.001^{***}$ \\
$p$ value & 0.00 & 0.00 & 0.00 & 0.00 & 0.00 & 0.13 & 0.00 & 0.00 \\
\bottomrule
\end{tabular}
}
\end{table}

\subsection*{Japanese firms}

\begin{figure}[tb]
\begin{minipage}{0.48\textwidth}
    \centering
    \includegraphics[width=\textwidth]{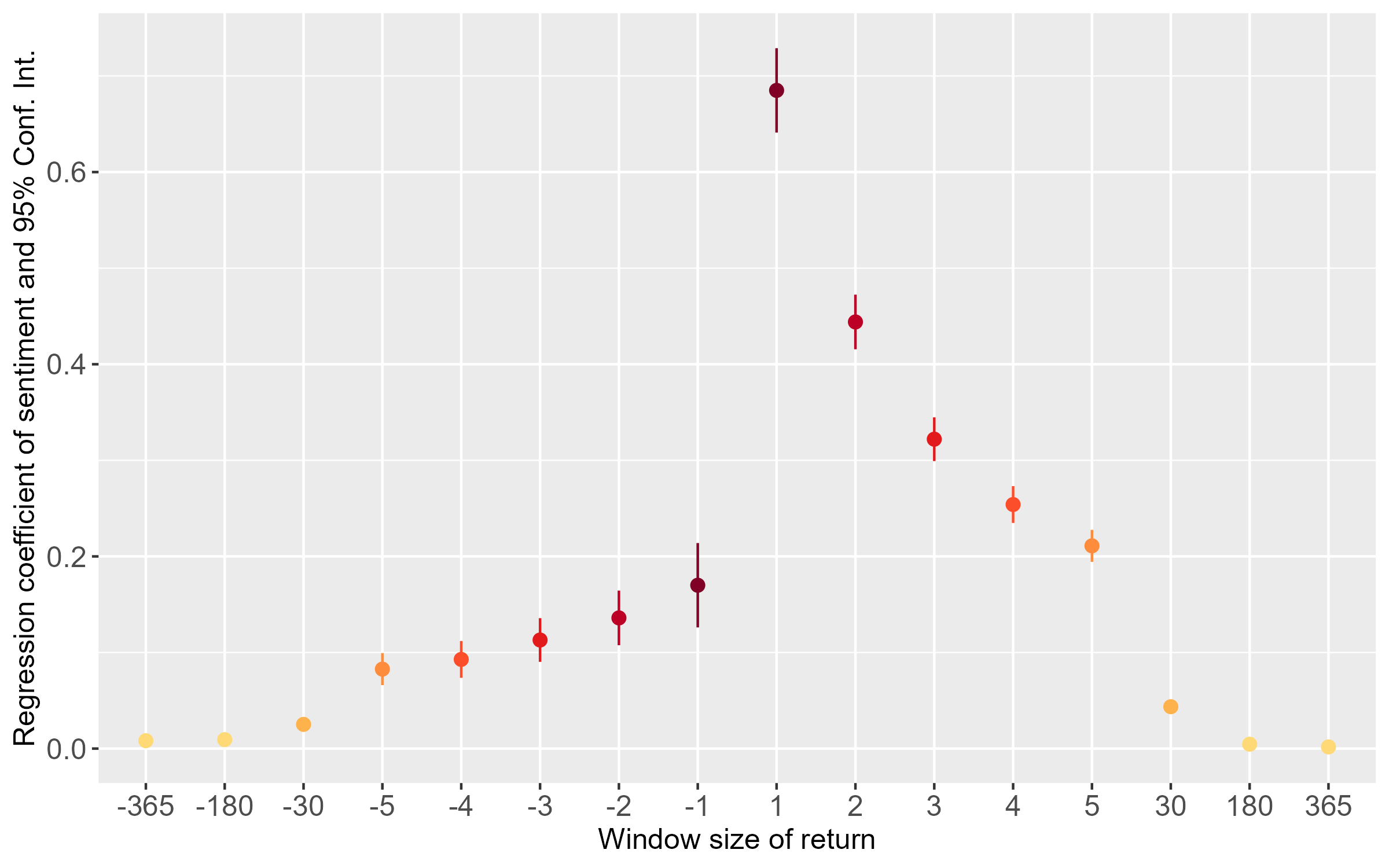} 
    \caption*{(A) Coefficient estimates of positive sentiment} 
\end{minipage}
\hfill
\begin{minipage}{0.48\textwidth}
    \centering
    \includegraphics[width=\textwidth]{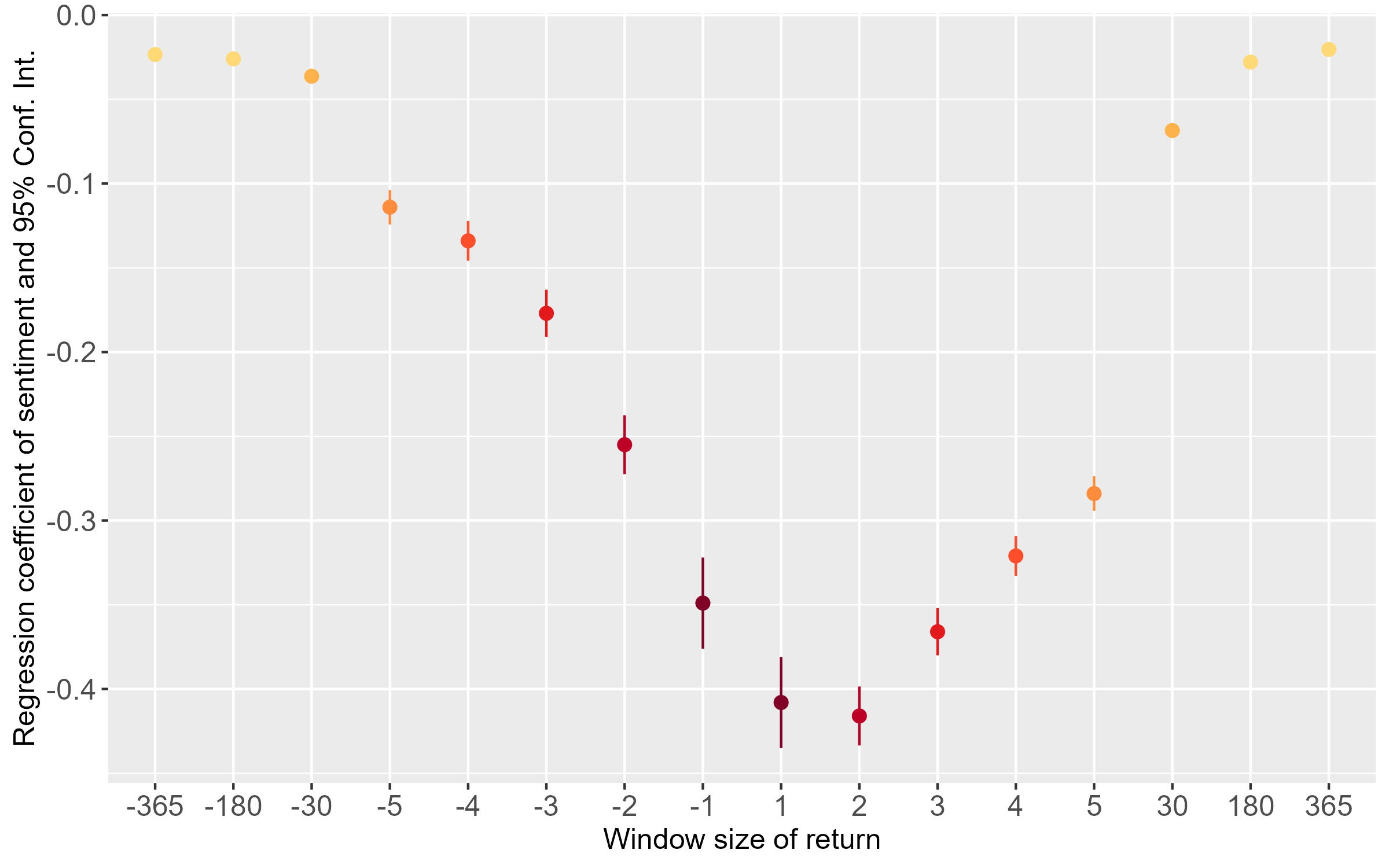}
    \caption*{(B) Coefficient estimates of negative sentiment} 
\end{minipage}
\caption{Coefficient estimates for the positive (panel (A)) and negative (panel (B)) sentiment indexes of news articles about \underline{Japanese firms} on percentage changes of their \underline{own} stock prices for different time windows. When the value of the horizontal axis is negative and $-w$, the dot above the value indicates the point estimate of the pre-news effect for time window $w$ ($\beta^{\mathit{pre}}_w$). When it is positive and $w$, the dot indicates the post-news effect for time window $w$ ($\beta^{\mathit{post}}_w$). The color of the dots for time windows $w$ and $-w$ is set to be the same so that the post- and pre-news effects can be easily compared. The confidence interval at the 5\% level associated with each point estimate (dot) is shown by a vertical segment.}
\label{fig:reg_japan_own}
\end{figure}

We further use another sample that focuses on Japanese listed firms. The results for the own-firm coefficients of positive and negative news sentiment for Japanese firms' own stock prices shown in Figure \ref{fig:reg_japan_own} are qualitatively and quantitatively similar to those using the sample of firms across the world shown in Figure \ref{fig:reg_global_own}. The detailed regression results are shown in SI Tables 8 for positive and 9 for negative, respectively. One subtle difference is that in Figure \ref{fig:reg_japan_own} for Japan, the post-news coefficient for negative news sentiment using the time window of 2 days is larger in absolute terms than that using the time window of 1 day, while Figure \ref{fig:reg_global_own} for firms in the world shows the opposite. This result suggests that the market reaction to negative news in Japan is slower than that to positive news in Japan and to positive and negative news in the world.   

In addition, we examine coefficients for news sentiment about Japanese firms on stock prices of their suppliers and clients. Detailed regression results for suppliers are reported in SI Tables 10 and 11, and those for clients are reported in SI Tables 12 for positive and 13 for negative, respectively. 

\begin{figure}[tb]
\begin{minipage}{0.48\textwidth}
    \centering
    \includegraphics[width=\textwidth]{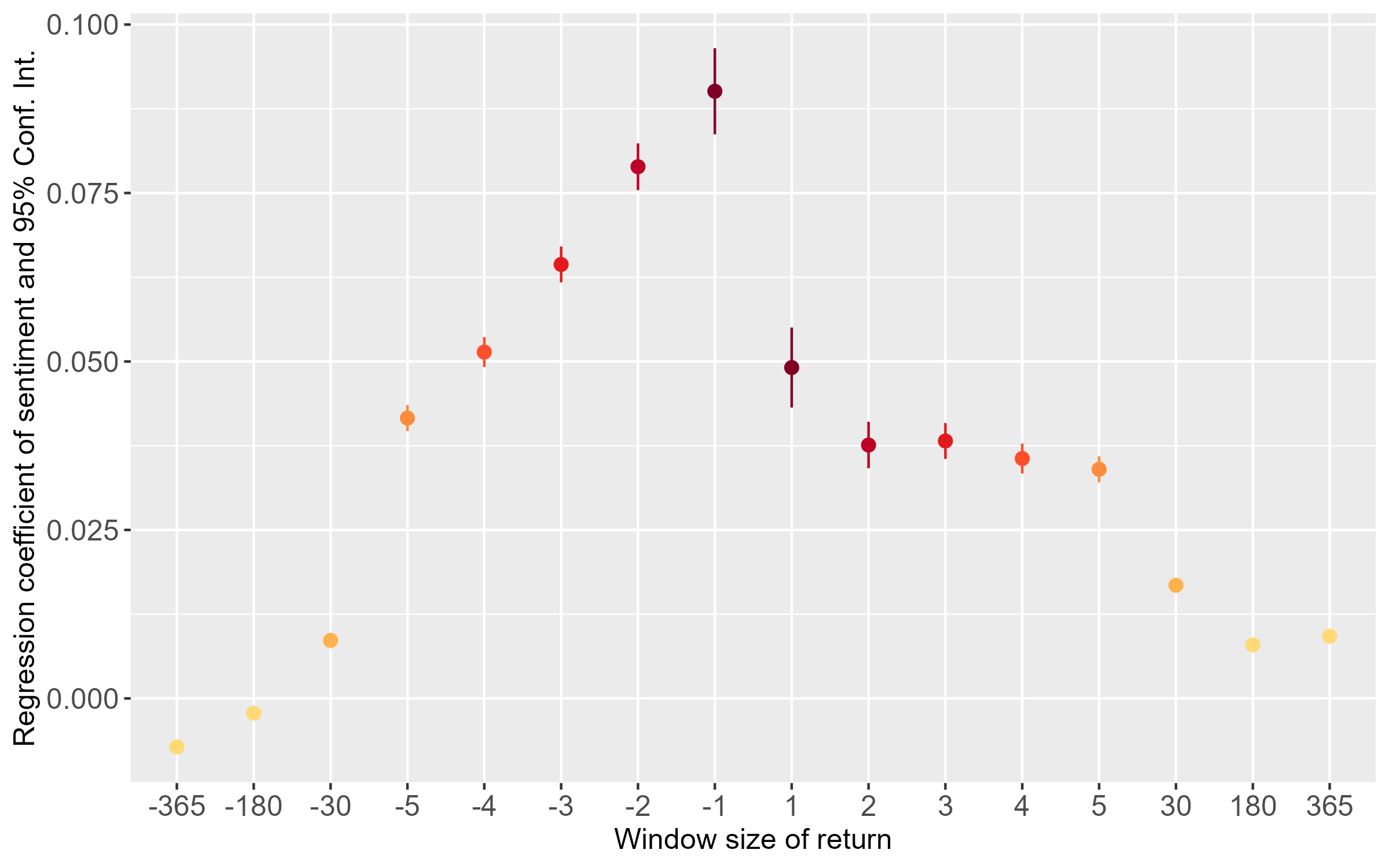} 
    \caption*{(A) Coefficient estimates of positive sentiment} 
\end{minipage}
\hfill
\begin{minipage}{0.48\textwidth}
    \centering
    \includegraphics[width=\textwidth]{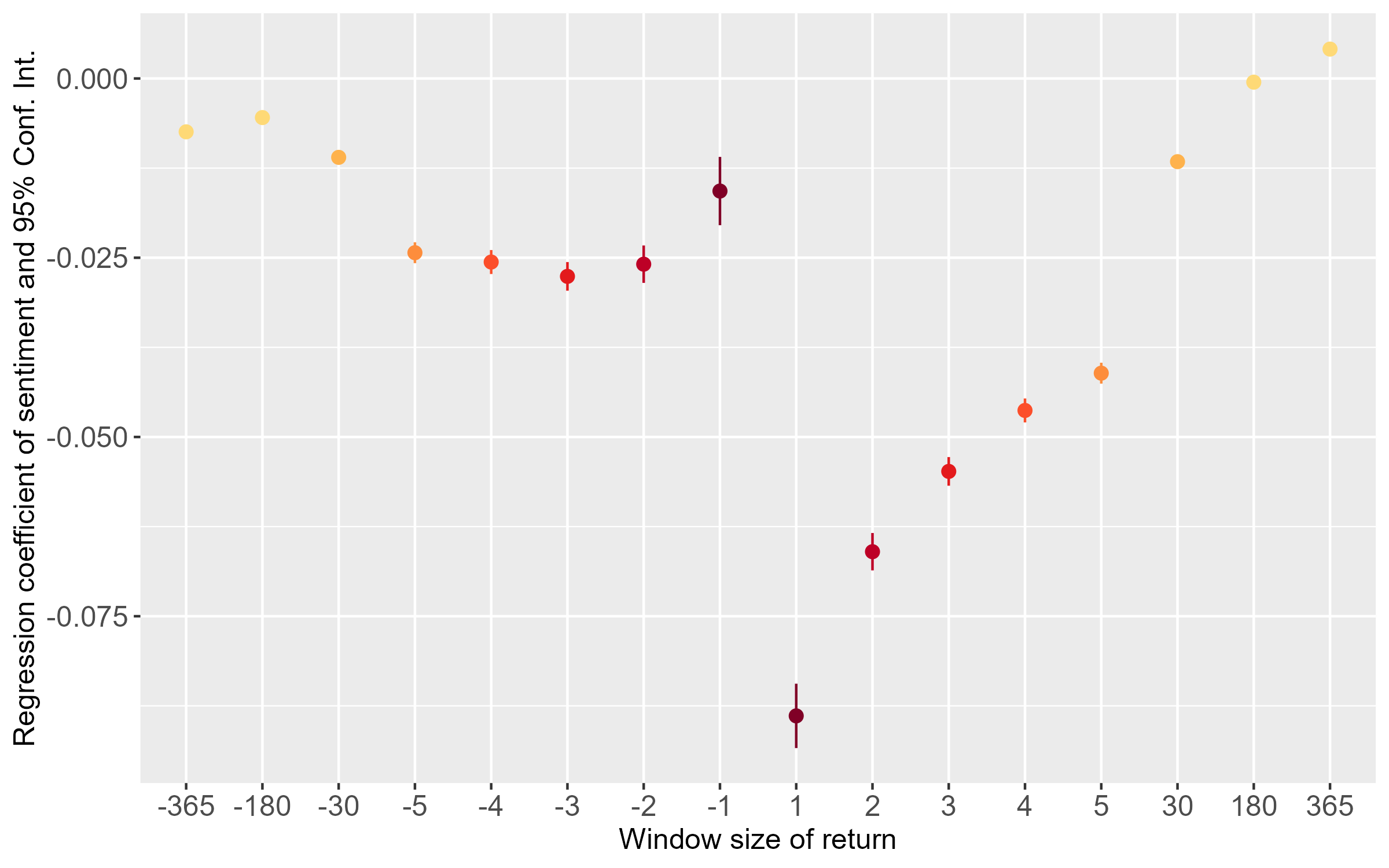}
    \caption*{(B) Coefficient estimates of negative sentiment} 
\end{minipage}
\caption{Coefficient estimates for the positive (panel (A)) and negative (panel (B)) sentiment indexes of news articles about \underline{Japanese firms} on percentage changes of their \underline{suppliers}' stock prices for different time windows. When the value of the horizontal axis is negative and $-w$, the dot above the value indicates the point estimate of the pre-news effect for time window $w$ ($\beta^{\mathit{pre}}_w$). When it is positive and $w$, the dot indicates the post-news effect for time window $w$ ($\beta^{\mathit{post}}_w$). The color of the dots for time windows $w$ and $-w$ is set to be the same so that the post- and pre-news effects can be easily compared. The confidence interval at the 5\% level associated with each point estimate (dot) is shown by a vertical segment.}
\label{fig:reg_japan_supplier}
\end{figure}

The coefficient estimates for positive and negative news sentiment in relation to suppliers' stock prices are shown in Panels (A) and (B) of Figure \ref{fig:reg_japan_supplier}, respectively, revealing three notable differences from the corresponding global results shown in Figure \ref{fig:reg_global_supplier}. First, the pre-news coefficients, which are larger in magnitude prior to disclosure, are generally larger for Japanese firms than for firms in the global sample. For example, for the one- and two-day time windows, the pre-news coefficients for positive news on suppliers are 0.0090 and 0.0079 in Japan, compared with 0.0029 and 0.0017 in the global sample. Similarly, for the same time windows, the post-news coefficients for negative news sentiment are -0.0089 and -0.0066 in Japan, compared with -0.0046 and -0.0025 in the global sample.

Second, the post-news coefficient for positive news sentiment is smaller than its pre-news coefficient (Panel (A) of Figure \ref{fig:reg_japan_supplier}), although the post-news coefficient is always larger in size than the pre-news coefficient in the global sample. This result implies that positive information about a firm is incorporated into prices prior to disclosure. However, once the positive information is disclosed in a news article, the magnitude of the coefficients for suppliers is smaller than that in the pre-news period. In other words, for Japanese firms, the difference between the post- and pre-news effects on suppliers is negative, indicating that the pre-news effect exceeds the post-news effect.

In contrast to the results for positive news, the response to negative news in Japan exhibits the same qualitative pattern as that observed in the global sample. Specifically, for negative news, the post-news effect is larger than the pre-news effect for Japanese firms, as is also the case for firms worldwide
(Panel (B) of Figure \ref{fig:reg_japan_supplier} and Panel (B) of Figure \ref{fig:reg_global_supplier}).

Within this common pattern in negative cases between the entire world and Japan, however, there are differences across time windows. In particular, for Japanese firms, the pre-news effect of negative news for the one-day time window is smaller in absolute value than that for longer windows of two to five days, whereas the global results show an increase in the absolute value of the pre-news effect as the window size decreases.

\begin{figure}[tb]
\begin{minipage}{0.48\textwidth}
    \centering
    \includegraphics[width=\textwidth]{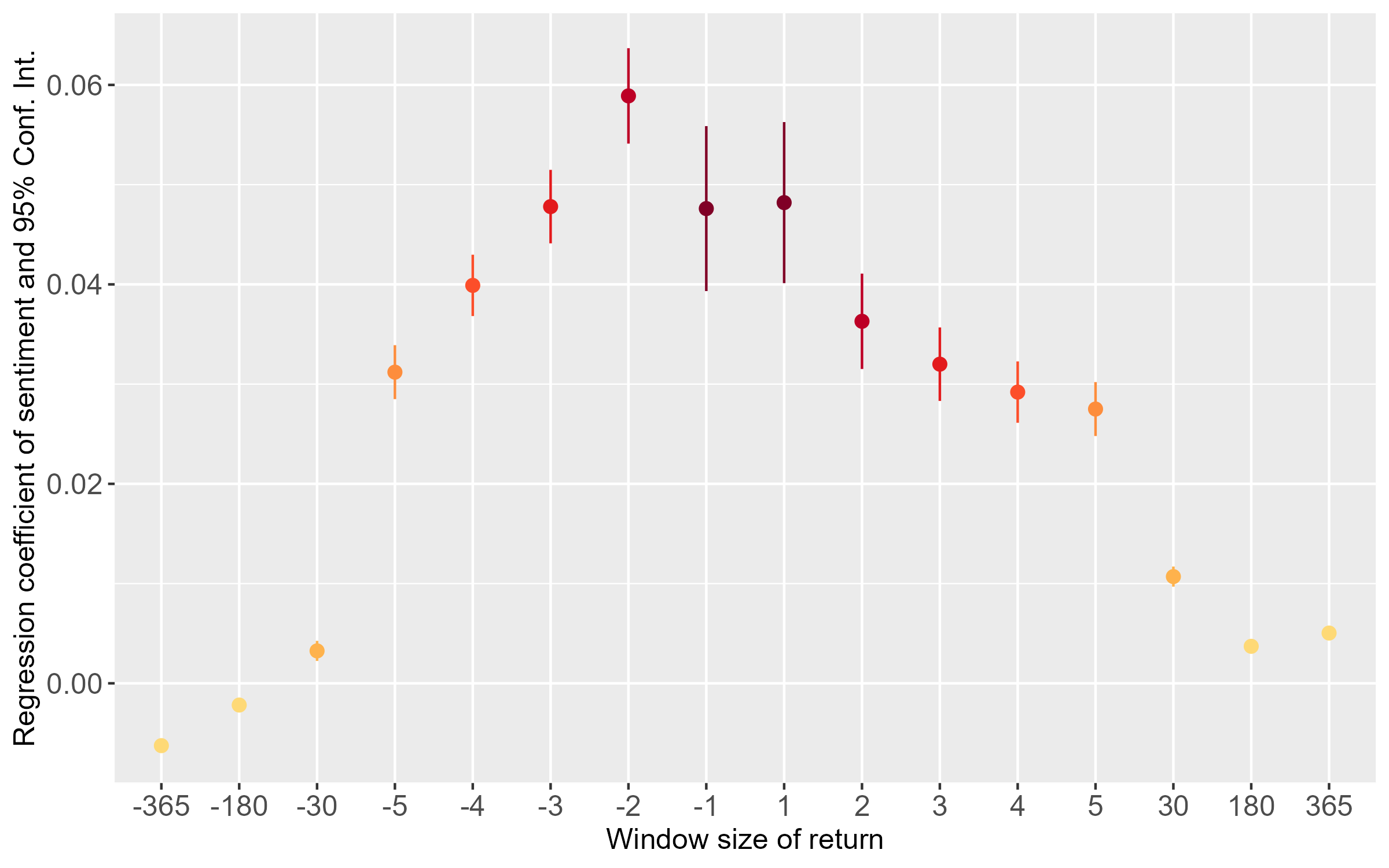} 
    \caption*{(A) Coefficient estimates of positive sentiment} 
\end{minipage}
\hfill
\begin{minipage}{0.48\textwidth}
    \centering
    \includegraphics[width=\textwidth]{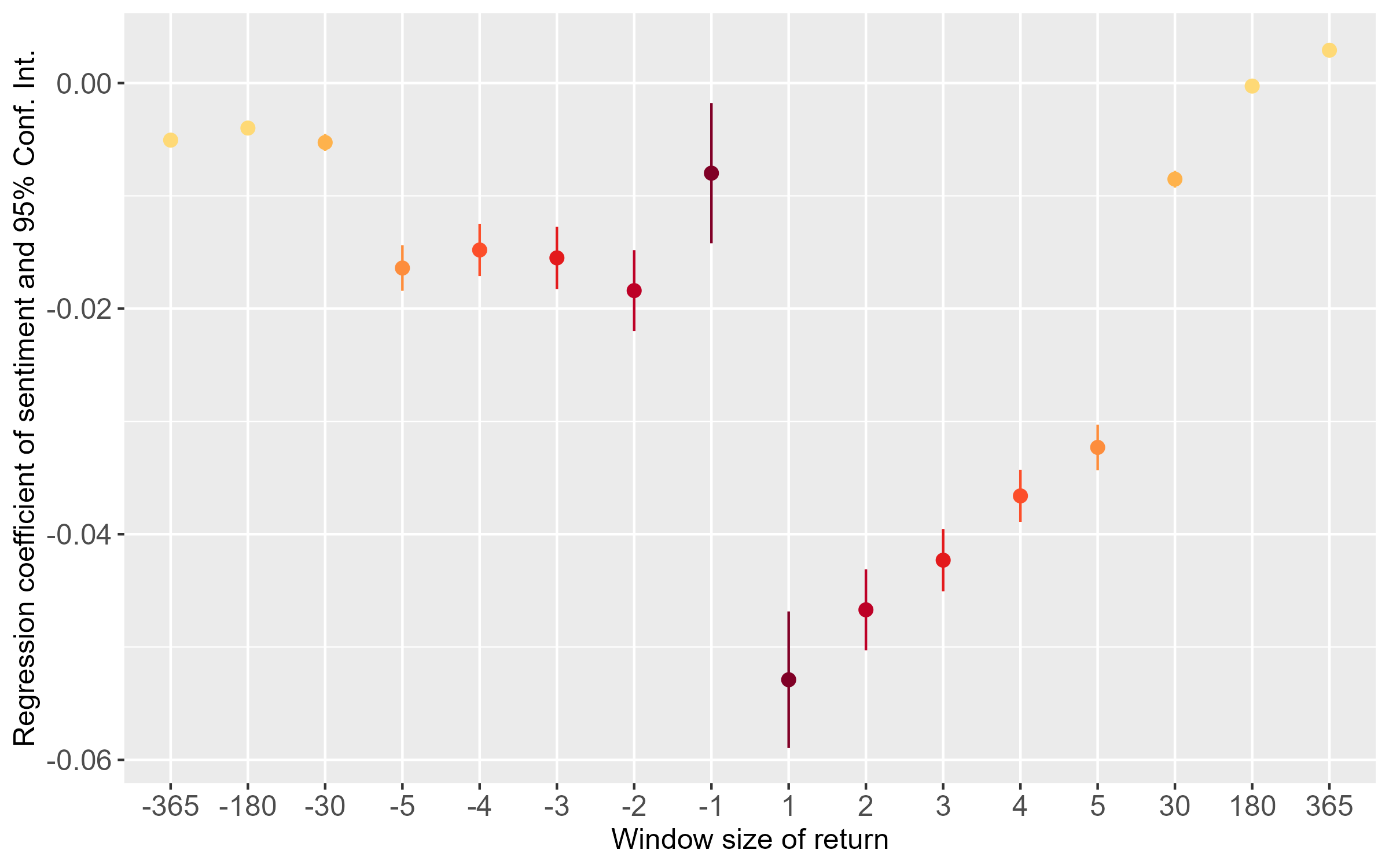}
    \caption*{(B) Coefficient estimates of negative sentiment} 
\end{minipage}
\caption{Coefficient estimates for the positive (panel (A)) and negative (panel (B)) sentiment indexes of news articles about \underline{Japanese firms} on percentage changes of their \underline{clients}' stock prices for different time windows. When the value of the horizontal axis is negative and $-w$, the dot above the value indicates the point estimate of the pre-news effect for time window $w$ ($\beta^{\mathit{pre}}_w$). When it is positive and $w$, the dot indicates the post-news effect for time window $w$ ($\beta^{\mathit{post}}_w$). The color of the dots for time windows $w$ and $-w$ is set to be the same so that the post- and pre-news effects can be easily compared. The confidence interval at the 5\% level associated with each point estimate (dot) is shown by a vertical segment.}
\label{fig:reg_japan_client}
\end{figure}

We further examine coefficients for news sentiment on client firms and present the results in Figure \ref{fig:reg_japan_client}. Detailed regression results are reported in SI Tables 12 for positive and 13 for negative, respectively. Compared with the corresponding results for firms across the world shown in Figure \ref{fig:reg_global_client}, we find that the same three differences observed for suppliers also apply to clients. First, the average magnitude of the effects is generally larger for Japanese firms than for firms in the global sample. Second, for positive news, the difference between the post- and pre-news effects on clients is negative in Japan. Third, for negative news, the pre-news effect for the one-day time window is smaller than that for longer time windows.

\section{Discussion}

In summary, we find that positive and negative news about firms are associated with stock price changes of their suppliers and clients before and after the disclosure of the information and the pattern of association differs between the world and Japan, which is observed by large scale data of news, stock prices, and supply chains. Additionally, we find that stronger pre-news stock price responses are observed for positive news in the Japanese sample.

As noted in the Introduction, we adopt a reduced-form perspective in the sense that we do not model investors’ information processing or decision-making. With this focus on price-based responses in mind, we turn to the interpretation of differences between the global and Japanese samples. The differences between the global and Japanese samples do not imply that different databases mechanically generate different results. Rather, the results consistently show that stock price responses suppliers' and clients' news in both samples, while the timing and relative magnitude of association differ across global and Japanese contexts. The FactSet database captures major supply chain links disclosed in financial statements, whereas the TSR database identifies economically important trading partners through firm-level surveys. These databases therefore provide qualitatively different views of supply chain relationships. The specific context of Japan may potentially contribute to the observed differences. Supply chain relationships in Japan are often discussed using the term keiretsu~\cite{Aoki1988keiretsu, McGuire2009keiretsu}. Such institutional difference may possibly affect the decision making of investors. The present analysis does not identify whether, or how, such institutional features contribute to the observed patterns. Other factors, such as historical shocks to Japanese supply chains due to disasters, such as the Great East Japan earthquake in 2011, may also play a role. Again, identifying the difference behind supply chains is beyond the scope of this study and remains an important direction for future research.

We examine the stock price responses, i.e., market valuation. While other market indicators such as trading volume or volatility capture important aspects of market activity and uncertainty, they represent different dimensions of market behavior. Analyses of news spillovers in trading volume or volatility would require a different empirical design and are therefore not pursued in this study.




\section{Conclusion}

We find that news sentiment, both positive and negative, is associated with systematically higher and lower change rates of stock prices of firms mentioned in the news in the pre- and post-news period. News sentiment is also associated with stock price responses of the firms’ suppliers and clients in both the pre-news and post-news periods, consistent with systematic valuation comovements along supply-chain relationships observed in other studies. In addition, we generally find that post-news effects on stock prices of the mentioned firms and their suppliers and clients are larger than pre-news effects. However, for Japanese firms, post-news effects on suppliers and clients can be smaller than pre-news effects, in contrast to the pattern observed for firms across the world.

The important opening question not addressed by this study is the significant heterogeneity of supply chain relationships, which can lead to varying impacts of news sentiment. For example, some suppliers are easily replaceable, while others are not; some trade relationships are long-term, potentially reflecting trust or sunk costs; additional links, such as shareholding or dual board membership, may exist; and the recognition of supply chain relationships by investors can vary. Examining the heterogeneity presents an important direction for future research.

\section*{Data availability}
The data used in this study are obtained from commercial databases, including FactSet Research Systems Inc., Tokyo Shoko Research, Ltd., and London Stock Exchange Group Data \& Analytics (Refinitiv).  These data are subject to licensing restrictions and cannot be made publicly available by the authors. Researchers with access to the same commercial data sources under appropriate licenses can obtain the data directly from the respective providers and reproduce the analyses conducted in this study. The authors do not have special access privileges to these data beyond standard institutional licenses.
 
\bibliography{ref}

\section*{Acknowledgements}

The authors thank Wataru Souma and Takayuki Yumoto for valuable discussions and Yuito Nishi for his technical contribution.

\section*{Funding information}
This work was supported by JSPS KAKENHI Grant Numbers JP22K18533, JP23K20626, JP23K25520, JP25K01454, JST PRESTO Grant Number JPMJPR21R2, and the Asahi Glass Foundation.

\section*{Author contributions statement}

H.I conceived the study.
H.I and Y.T conducted the analyses, and wrote and reviewed the manuscript.

\clearpage

\end{document}